\documentclass[11pt]{article}
\usepackage{latexsym}
\usepackage{amsmath}
\usepackage{amsfonts}
\usepackage{mathrsfs}
\usepackage{dsfont}
\usepackage{enumerate}
\usepackage{verbatim}
\usepackage{hyperref}
\usepackage{tikz}
\usepackage{mathtools}
\usepackage{simplewick}
\usepackage{float}
\usepackage{caption}
\usepackage{slashed}
\usepackage[title]{appendix}
\usepackage{subcaption}
\usepackage{graphicx}

 \csname
@addtoreset\endcsname{equation}{section}
\usepackage{amsmath,amssymb}
\usepackage{graphicx}
\usepackage{color}

\begin{document}

\def\be{\begin{equation}}
\def\ee{\end{equation}}
\def\bea{\begin{eqnarray}}
\def\eea{\end{eqnarray}}
\def\nn{\nonumber}

\renewcommand{\thefootnote}{\fnsymbol{footnote}}
\renewcommand*{\thefootnote}{\fnsymbol{footnote}}

\begin{flushright}

\end{flushright}

\vspace{40pt}

\begin{center}

{\Large\sc Noncommutative spacetime geometry and one-loop effects in primordial cosmology}

\vspace{50pt}

{\sc Hai Siong Tan}

\vspace{15pt}
{\sl\small Division of Physics and Applied Physics,
School of Physical and Mathematical Sciences, \\
Nanyang Technological University,\\
21 Nanyang Link, Singapore 637371}

\vspace{15pt}

\vspace{70pt} {\sc\large Abstract}\end{center}
We study the effect of noncommutative spacetime geometry on one-loop corrections
to the primordial curvature two-point function, arising from various forms of massless spectator matter fields interacting gravitationally 
with the inflaton. After deforming the algebra of functions on the inflationary background to a spatially noncommutative one, 
we find that this induces momentum-dependent corrections to one-loop terms which imply that the 
vacuum fluctuations of the energy-momentum tensor source that of the curvature fluctuation
even for distances beyond horizon scales. The one-loop corrections break spatial isotropy by being functions
of the noncommutative parameters lying in the tranverse plane while reducing smoothly 
to the commutative limit. This furnishes an example of how UV/IR mixing manifests itself in the context of 
noncommutative field theories defined on inflationary backgrounds, and demonstrates how in principle,
the primordial spectrum could carry a signature of nonlocality and anisotropy in the setting of noncommutative 
spacetime geometry.

\newpage

\tableofcontents

\renewcommand*{\thefootnote}{\arabic{footnote}}
\setcounter{footnote}{0}

\section{Introduction}

For the primordial universe where quantum gravitational effects were very strong, the notion of a classical spacetime continuum should break down at some point and stringy effects could be dominant in such an era. One of the simplest models of a departure from the continuum notion is that of `discretized' spacetime \cite{Snyder,Yang} that follows from a noncommutative structure like that of the canonical quantization bracket in quantum mechanics. In string theory, noncommutative spacetime geometry can emerge in various settings, a few examples being: (i) a collection of D0 branes in the presence of external RR four-form flux expanding into a noncommutative $S^2$ \cite{Myers}, (ii) vertex operators of open strings following an associative but noncommutative Moyal star product in the presence of a large NS-NS B-field $B$ in the zero slope ($\alpha' \rightarrow 0$) limit, with the target space coordinates obeying a noncommutative bracket \cite{Witten} 
$$
[x^i, x^j ] \sim i\theta^{ij} \sim i (1/B)^{ij}.
$$
 Motivated by such considerations, there has been interesting works that have sprung up over the past twenty years studying issues of how noncommutative geometry could potentially manifest itself in primordial cosmology especially in the context of inflationary dynamics  \cite{Henry,Shiu,Buchel,Alexander}. For example as explained in \cite{Shiu}, noncommutative inflation can arise if the inflaton field originates from the open string sector such as being the modulus of the endpoint of open strings, while the gravitational background remains unaffected by the noncommutativity.  
At the phenomenological level, one typically proceeds to develop an effective description via a study of how noncommutative deformations of field theories on cosmological spacetime backgrounds should work, and their implications for observables such as correlation functions.

For the one-loop runnings of the primordial spectrum in the usual `commutative' setting, they were first computed by Weinberg  \cite{Weinberg1} via the Schwinger-Keldysh (or `in-in') formalism with the results generalized and refined by several others, eg. \cite{Senatore,Eugene,Chai,Tanaka,Giddings}. They could arise in various manners, such as self-interaction of the inflaton \cite{Seery1}, graviton \cite{Feng}, etc. For our purpose here, we will be interested in loop corrections due to external matter fields first considered in \cite{Weinberg1}.
From the phenomenological point of view, there is unfortunately little hope of observing the one-loop correction which is roughly speaking suppressed by a factor of $\sim \epsilon \left( \frac{H}{M_{Pl}} \right)^2$ where $\epsilon$ is the slow-roll parameter, $H$ is the Hubble parameter during inflation and $M_{Pl}$ the Planck mass, relative to the tree-level two-point function.
From our current observations of the CMB, this factor should be less than an order of $10^{-12}$. Yet as Weinberg first explained in \cite{Weinberg1}, there are good reasons to study these corrections especially for external matter fields that are present prior to reheating which parametrically dominate over the corresponding one-loop effect from gravitons by virtue of their numbers. At least in principle, it would be interesting to know how these effects distinguish between different types of matter fields, if they could grow large with time and induce deviations of the primordial spectrum from scale invariance. 

Refining the results in \cite{Weinberg1},  Senatore and Zaldarriaga \cite{Senatore} showed that the one-loop correction, apart from 
a renormalization constant term, takes the form of a logarithmic running $\log \left(  \frac{H}{\mu} \right) $ where $\mu$ denotes the renormalization scale. This matches well with a typical logarithmic term appearing in scattering amplitudes $\log \left(  \frac{E}{\mu} \right)$
where $E$ is the invariant collision energy. Such a one-loop term is compatible with eternal inflation, invariant under the rescaling symmetry (i.e. $a\rightarrow \lambda a, x \rightarrow x/\lambda$ where $a$ is the scale factor ) of the FRW metric and preserves scale invariance. (Previously in \cite{Weinberg1} and other preceding works such as \cite{Eugene}, the logarithmic running is a function of the external momentum $\log (k/\mu )$.) As mentioned in \cite{Senatore}, the result $\log \left(  \frac{H}{\mu} \right)$ is compatible with what we expect from a weakly coupled theory for which the one-loop logarithmic term would be small for a renormalization scale $\mu$ that is near the energy scale of the process. (In this case, for example, some order-of-magnitude theoretical estimates yield $H \sim 10^{-5} M_{Pl}, \mu \sim 10^{-3} M_{Pl}$ if we renormalize it at GUT scale.)

In this paper, we consider one-loop corrections to primordial spectrum arising from massless matter fields that interact purely gravitationally 
with the inflaton coupled with the backdrop that they reside on a noncommutative spacetime, in this case, a slow-roll inflationary background. We find that the noncommutative deformation of the operator algebra introduces momentum dependence into
the one-loop correction, which implies that the vacuum fluctuation of the matter field source that of the curvature fluctuation
even for distances beyond the horizon scale - a striking feature of non-locality induced by the noncommutative spacetime geometry. 
Further, the one-loop corrections break spatial isotropy by being functions
of the noncommutative parameters lying in the tranverse plane\footnote{Defining $\theta^i = \frac{1}{2}\epsilon^{ijk} \theta_{jk}$, the one-loop corrections
depend on the magnitude of the components of $\vec{\theta}$ transverse to the momentum vector of the curvature fluctuations in momentum space.} 
while reducing smoothly 
to the commutative limit. This furnishes an example of a UV/IR mixing process transmitting the noncommutativity scale to superhorizon scale
in the context of noncommutative field theories defined on an inflationary background.  The momentum-dependent 
one-loop correction demonstrates how in principle,
the primordial spectrum could carry a signature of nonlocality in the setting of noncommutative 
spacetime geometry.  

We perform this calculation in the comoving curvature gauge (in which inflaton fluctuations are set to zero) for the cases of massless scalars, fermions and U(1) gauge fields. In \cite{Chai}, the trilinear interaction Hamiltonian describing their gravitational interaction with the curvature mode was derived. It is straightforward to write down their noncommutative generalizations (for the fermions and complex scalars, it turns out that there is a one-parameter family of deformations) and deriving the one-loop correction amounts to computing various four-point functions
that could be performed by integrating products of Wightman functions. Graphically, we can represent the one-loop term as a sum of planar
and non-planar Feynman graphs as depicted in Figure 1.

\begin{figure}[h]
\centering
\includegraphics[width=170mm]{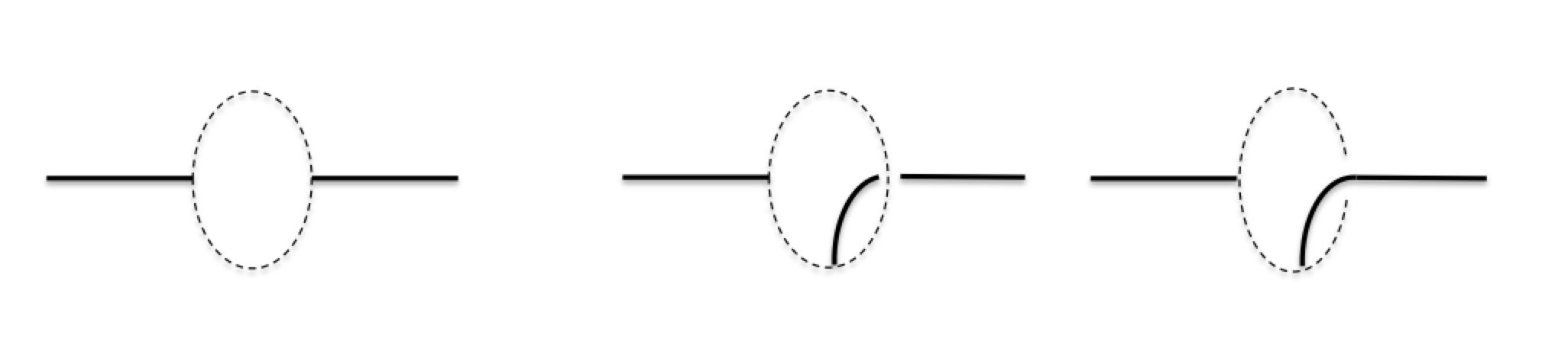}
\vspace{10pt}
\caption{In the set of Feynman diagrams above, bold lines represent the primordial curvature $\zeta$ while
the dotted lines represent some matter fields. In this paper, we compute these one-loop diagrams for massless scalars, fermions and U(1)
gauge fields.
The leftmost planar graph is the usual 
correlation function obtained in the absence of noncommutative deformation, whereas the non-planar graphs 
contain phase factors that accompany the vertices. They are schematically of the form 
$e^{\frac{i}{2} \sum_{i,j} I_{ij} \theta_{\mu \nu} k^{\mu}_i k^{\nu}_j  }$ where $I_{ij}$ is a matrix equal to $\pm 1$ depending on whether the line $j$ (with momentum $k_j$) crosses line $i$ (with momentum $k_i$) from the right or left respectively, being zero in the absence of crossing, and $\theta$ is the noncommutative constant matrix-valued parameter. }
\label{fig1}
\end{figure}

In this work, we compute the various one-loop terms (both planar and non-planar graphs) using both independent methods of
dimensional regularization and cutoff regularization, making sure that in all cases we obtain the 
same finite quantum correction. We assume purely a spatial and constant noncommutativity parameter without assuming any specific stringy realization, 
leaving more elaborate
generalizations for future work. Our results add to previous work in literature which explores 
the role of noncommutative geometry in inflationary dynamics. In \cite{Akofor1}, it was shown that assumption of noncommutative spacetime geometry parametrized by $\theta^{0i}$ (i.e. space-time noncommutativity) leads to a deformation of the primordial
power spectrum and in \cite{Akofor2}, a bound ($H\theta < 0.01$ Mpc)on the noncommutative parameter was developed via comparison with WMAP5 data. Further in \cite{Nautiyal}, the noncommutative modification of the non-gaussianity was computed. Similar to these results,
we find that after imposing noncommutative geometry, the one-loop correction breaks spatial isotropy and represents a non-local 
coupling between the noncommutative background and fluctuation modes.

The plan of our paper goes as follows. In Section 2, we begin with a review of the derivation of 
one-loop correction in the usual commutative case for scalars, fermions and vector fields via two different
regularization procedures. Incidentally, there are some subtleties in the Schwinger-Keldysh formalism associated with 
an $i\epsilon$-prescription in time-ordering which have affected the accuracy of some of the previous 
results reported in literature, so we also take the opportunity to 
refine the accuracy of
some of the one-loop correction terms 
in the usual commutative case. In Section 3, we discuss 
some aspects of quantum field theories on noncommutative spacetime, in particular the noncommutative deformations of
spectator matter fields residing on an inflationary spacetime background. We derive the general form of the phase factor that one
could associate with the interaction vertices. Section 4 presents the main results of our paper - the derivation of various non-planar
one-loop corrections due to various matter fields together with some comments on their physical interpretations. Finally, 
we end off in Section 5 with some concluding remarks.

\section{One-Loop corrections to primordial spectrum from matter fields }

\subsection{Some preliminaries}

In \cite{Weinberg1}, Weinberg formulated the framework for computing loop corrections to cosmological 
correlation functions due to matter fields, and this was further elaborated nicely in \cite{Chai}. 
The classical background is that of standard slow-roll inflation coupled to external matter fields. Loop corrections
arising from the matter fields are typically enhanced by degeneracy factors (corresponding to the potentially large number of types of fields) and hence are generically more significant than graviton and inflaton loops. 

We consider the (minimal) coupling of external matter fields to the gravitational action equipped with a slow-roll inflaton potential, with the Lagrangian 
\be
\mathcal{L}_g = -\frac{1}{2}\sqrt{-g} \left( \frac{1}{8\pi G}R + \partial^\mu \phi \partial_\mu \phi +  2V(\phi) \right) + \mathcal{L}_{matter},
\ee
where $\phi,\, V(\phi)$ refer to the inflaton and its potential respectively.
In subsequent sections, we consider three basic massless matter fields: scalars ($\chi$), fermions ($\Psi$) and abelian gauge fields ($A_{\mu}$), with $\mathcal{L}_{matter}$ being
\be
\mathcal{L}_s = -\frac{1}{2} \sqrt{-g} \partial^\mu \chi \partial_\mu \chi,
\qquad
\mathcal{L}_f = -\frac{1}{2} \sqrt{-g} \left[
\overline{\Psi} \gamma^\alpha D_\alpha \Psi - (D_\alpha \overline{\Psi} ) \gamma^\alpha \Psi \right],
\qquad
\mathcal{L}_g = -\frac{1}{4} \sqrt{-g}   F^{\mu \nu} F_{\mu \nu},
\ee
where $D_\alpha \Psi = (\partial_\alpha + \frac{1}{4} {\omega_\alpha}^{mn} \sigma_{mn} ) \Psi$ is the covariant derivative for the Dirac spinor with $\omega$ being the spin connection and $\sigma_{mn} = -\frac{1}{2} [\gamma_m, \gamma_n]$, with $\gamma_m$ being the tangent space Dirac matrices. In cosmological perturbation theory, one expands the action to arbitrary order in fluctuations of the metric, inflaton and various matter fields. A convenient approach is to work in the ADM formalism with some judicious choice of gauge. In the following we are interested in computing one-loop corrections to the primordial spectrum $\langle \zeta \zeta \rangle$, with $\zeta$ defined in the following spatial metric decomposition 
\be
g_{ij} = a^2 e^{2\zeta} \left[ e^{\gamma} \right]_{ij}, 
\ee
where $a(t)$ is the scale factor, $\gamma_{ij}$ is the (traceless and tranverse) gravitational wave amplitude.
and we have picked a gauge in which the inflaton does not fluctuate. The other metric components are parametrized by 
$$
g_{00} = -N^2 + g_{ij} N^i N^j, \qquad 
g_{i0} = g_{ij} N^j,
$$
where $N, N_i$ are the lapse and shift auxiliary fields that can be solved by the Hamiltonian constraint equations. 
Expanding all matter and gravitational fluctuations to arbitrary orders is tedious yet straightforward in principle, a process that can be simplified using suitable field redefinitions and classical equations of motion. 
For our present purpose of computing one-loop corrections, we require the explicit form of cubic interaction terms which are fortunately derived in \cite{Chai} for various types of matter fields. These trilinear interaction Hamiltonians can be written compactly in terms of components of the energy-momentum tensor as follows \cite{Chai}
\be
\label{Hint}
H_{int} (t) = - \int d^3x\, \epsilon H a^5 ( T^{00} + a^2 T^{ii} ) \nabla^{-2} \dot{\zeta},
\ee
where $\epsilon = -\dot{H}/H$ is the slow-roll parameter. 
For massless scalars, Dirac fermions and U(1) gauge fields, the interaction Hamiltonian \eqref{Hint} reads
\bea
\label{H1}
H_{\zeta \chi \chi} &=& -\int d^3x\, \epsilon H a^5 \left(  2 \dot{\chi}^2 \nabla^{-2} \dot{\zeta} \right), \cr
\label{H2}
H_{\zeta \Psi \Psi} &=& -\int d^3x\, 2 \epsilon H a^5 \left(   
\overline{\Psi} \gamma^0 \dot{\Psi} - \dot{\overline{\Psi}}\gamma^0 \Psi
\right)\nabla^{-2} \dot{\zeta}  \cr
\label{H3}
H_{\zeta A A} &=& -\int d^3 x\,\epsilon H a^5 \left(
\frac{1}{a^2}\dot{A}^2_i + \frac{1}{2a^4} ( \partial_i A_j - \partial_j A_i )^2
\right) \nabla^{-2} \dot{\zeta} 
\eea
In the Schwinger-Keldysh (or `in-in') formalism, we compute the expectation values of some product of field operators $Q(t)$ evaluated at some common time $t$ in the interaction picture, with the prescription 
\be
\label{SK}
\langle \Omega | Q (t) | \Omega \rangle = 
\langle 0 | \left[  
\bar{T} \text{exp} \left( i \int^t_{-\infty_+} dt\, H_{int} (t) 
\right) 
\right] Q(t) 
\left[  
T \text{exp} \left( - i \int^t_{-\infty_-} dt\, H_{int} (t) 
\right) 
\right] |0 \rangle .
\ee
In \eqref{SK}, the vacuum state of the free theory $| 0 \rangle$ is obtained after projecting on the interacting vacuum state $| \Omega \rangle$ with an $i\epsilon$ presciption, with the infinities analytically continued as 
\be
\infty_{\pm} = \infty (1 \pm i\epsilon ),
\ee
with $\epsilon$ being a real and positive regulator. In the interaction picture, the field operators evolve via the free Hamiltonian and thus they are expanded in terms of modes which are solutions to the Mukhanov free field equations. The expectation value of $Q(t)$ is then computed in perturbation theory by expanding \eqref{SK} to the required order. In this work, we focus on loop corrections to the primordial spectrum and thus $Q(t) = \zeta (t) \zeta (t)$, evaluated at large $t \sim \infty$ or in terms of conformal time $\tau \sim  0$.

For our purpose of computing the one-loop corrections arising from the trilinear interaction Hamiltonian, we require second order terms in \eqref{SK} which in conformal time read
\bea
\label{oneloopSK}
\langle \zeta \zeta \rangle_{1-\text{loop}} &=&
-2 \text{Re} \int^0_{-\infty_+} d\tau_2\, \int^{\tau_2}_{-\infty_+} d\tau_1\, \langle 0 | H_{\zeta M M}( \tau_1 ) H_{\zeta M M} (\tau_2) \zeta (0) |0\rangle  \cr
&&\qquad+ \int^0_{-\infty_-} d\tau_1 \int^0_{-\infty_+} d\tau_2\,\, \langle 0|  H_{\zeta MM} (\tau_1) \zeta^2 (0)  H_{\zeta MM} (\tau_2)  | 0 \rangle . 
\eea
On this note, we wish to point out that in the earlier works of \cite{Weinberg1} and \cite{Chai}, 
instead of \eqref{oneloopSK}, the following form for the second-order correction
was used to compute the one-loop correction. 
\be
\label{oneloop}
\langle Q \rangle = - \int^t_{-\infty} dt_2 \int^{t_2}_{-\infty} dt_1 \langle 0| 
\left[ H_{\zeta M M}( \tau_1 ), \left[ H_{\zeta M M}( \tau_2 ), Q \right] \right] | 0 \rangle ,
\ee
with the analytic continuation of $\infty (1\pm i\epsilon )$ adopted implicitly. It appears that the second line of \eqref{oneloopSK} would however be missing, which was pointed out in \cite{Eugene} and \cite{Senatore} to give rise to a different result for the one-loop correction (results were stated for the scalar field). In the following, we will use \eqref{oneloopSK} and find that the second line turns out to vanish in any case for the massless fermion and abelian gauge field, while it is opposite in sign and larger relative to the first line in \eqref{oneloopSK}, effectively rendering the scalar contribution to be opposite in sign to that of the fermion and gauge fields.

\subsection{On one-loop corrections from scalar, fermions and vector fields}

To proceed, we require the mode expansions of various fields : 
the scalars, fermions and gauge field operators can be expanded as
\bea
\label{m1}
\chi (\vec{x},t) &=& \int d^3 q \,\, e^{i\vec{q} \cdot \vec{x}}    \left[   \chi_q (t) a_{\vec{q}} + \chi^*_q (t) a^\dagger_{-\vec{q}} \right], \\
\label{m2}
\zeta (\vec{x},t) &=&  \int d^3 q \,\, e^{i\vec{q} \cdot \vec{x}}    \left[   \zeta_q (t) a_{\vec{q}} + \zeta^*_q (t) a^\dagger_{-\vec{q}} \right], \\
\label{m3}
A_i (\vec{x}, t) &=& \int d^3 q \, \sum_{\lambda}\, e^{i\vec{q} \cdot \vec{x} } \left[ A_q (t) e_i (\hat{q}, \lambda ) a_{\vec{q}, \lambda} + A^*_q (t) e^*_i (-\hat{q}, \lambda ) a^{\dagger}_{-\vec{q}, \lambda}  \right], \\
\label{m4}
\Psi (\vec{x}, t) &=& a^{-\frac{3}{2}} (t) \int d^3q\,\, \sum_s \, e^{i\vec{q} \cdot \vec{x}} 
\left[   U_{\vec{q}, s} (t) a_{\vec{q},s} + V_{-\vec{q}, s} (t) \beta^\dagger_{-\vec{q}, s}
\right], 
\eea
where $\lambda$ and $s$ label the polarization and spin basis states respectively.
The oscillators obey the usual (anti-)commutation relations and the modes satisfy the classical equations of motion. 
Substituting \eqref{m1} - \eqref{m3} into \eqref{oneloopSK}, in momentum space,
the one-loop correction can be expressed in the form
\be
\langle \zeta_{\vec{q}} \zeta_{-\vec{q}} \rangle_{1-\text{loop}} 
= \int d^3p d^3 p' \delta^3 (\vec{q} + \vec{p} + \vec{p'} ) 
\left[ -2\text{Re} \int^0_{-\infty_+} d\tau_2 \int^{\tau_2}_{-\infty_+} d\tau_1 \,\, Z_1 \mathcal{M} + 
\int^0_{-\infty_+} d\tau_1\, \int^0_{-\infty_-} d\tau_2\, Z_2 \mathcal{M} \right],
\ee
with $\mathcal{M}$ the four-point function of the matter fields (and their derivatives), and $Z_{1,2}$ being that of the curvature perturbation $\zeta$. In the following, as in \cite{Weinberg1} and \cite{Chai}, we will take the slow-roll and Hubble parameters to be slowly-varying and approximated as $H (\tau_1 ) \approx H (\tau_2) \approx H(\tau_q ), \epsilon (\tau_1) \approx \epsilon (\tau_2) \approx \epsilon (\tau_q)$, i.e effectively constants at the time of horizon exit. The time-integrals are, as explained in \cite{Weinberg1}, dominated by the time $\tau_q$ of horizon exit. The various unperturbed fields are rolling very slowly down the potential at $\tau_q$. At horizon exit, the scale factor is approximately of the form $e^{Ht}$. In \eqref{m1}-\eqref{m4}, the various wavefunctions with Bunch-Davies initial condition read
\bea
\label{modes}
\zeta_q (\tau) &=& \zeta^0_q e^{-iq\tau} (1+iq\tau ),  \qquad |\zeta^0_q|^2 = \frac{H^2 (\tau_q)}{2(2\pi)^3 \epsilon (\tau_q) q^3}, \cr
\chi_q (\tau) &=& \chi^0_q e^{-iq\tau} (1+iq\tau ),  \qquad |\chi^0_q|^2 = \frac{H^2 (\tau_q)}{2(2\pi)^3 q^3},
\cr
A_q (\tau)&=& A^0_q e^{-iq\tau}, \qquad |A^0_q|^2 = \frac{1}{2(2\pi)^3 q^3}, \cr
U_{\vec{q},s} (\tau) &=& U^0_{q,s} e^{-iq\tau}, \,\,\, 
V_{\vec{q},s} (\tau) = V^0_{q,s} e^{iq\tau}, \,\,\, 
\sum_s U^0_{q,s} \bar{U}^0_{q,s} = \sum_s V^0_{q,s} \bar{V}^0_{q,s} =  -\frac{i\gamma^\mu q_\mu}{2(2 \pi)^3 q},\,\, 
q = q^0. \nonumber \\
\eea
With these classical mode solutions, 
the 4-point function for the matter fields can be computed to read 
\bea
&& \textbf{Scalar} : 
\mathcal{M}_s = 4 \epsilon_1 \epsilon_2 H_1 H_2
a^2_1 a^2_2 (2\pi )^3 pp' e^{-i (p+p')(\tau_1 - \tau_2)},\cr \cr
&& \textbf{Fermions} : \mathcal{M}_f = 8 \epsilon_1 \epsilon_2 H_1 H_2
a^2_1 a^2_2 (2\pi )^3 (p-p')^2 
(1+ \hat{p} \cdot \hat{p}' )
 e^{-i (p+p')(\tau_1 - \tau_2)}, \cr \cr
\label{Mfields}
&& \textbf{U(1) gauge field}: \mathcal{M}_g = 2 \epsilon_1 \epsilon_2 H_1 H_2
a^2_1 a^2_2 (2\pi )^3 pp' (1+ \hat{p} \cdot \hat{p}' )^2  e^{-i (p+p')(\tau_1 - \tau_2)},
\eea
whereas those for the curvature perturbation read
\be
\label{Ztime}
Z_1 = \frac{H^2}{4a^2_1 a^2_2 \epsilon^2 (\tau_q ) (2\pi)^6 q^2} \left(  
e^{-iq (\tau_1 + \tau_2)} - e^{iq (\tau_2- \tau_1)}
\right), \qquad 
Z_2 = \frac{H^2}{4a^2_1 a^2_2 \epsilon^2 (\tau_q) (2\pi)^6 q^2} e^{iq (\tau_2- \tau_1)} . 
\ee
The time-integral is identical for all matter fields and can be easily performed.
\bea
\label{time1}
-2\text{Re} \int^0_{-\infty_+} d\tau_2\, \int^{\tau_2}_{-\infty_+} d\tau_1 \left( 
e^{-iq (\tau_1 + \tau_2) } - e^{iq (\tau_2 - \tau_1)} 
\right) e^{i(p+p')(\tau_2 - \tau_1)} &=& \frac{1}{q(p+p'+q)}, \\
\label{time2}
\int^0_{-\infty_+} d\tau_1\, \int^0_{-\infty_-} d\tau_2 
e^{iq(\tau_2 - \tau_1) + i(p+p')(\tau_2 - \tau_1)}
&=& \frac{1}{(p+p'+q)^2}. 
\eea
Assembling \eqref{Mfields} - \eqref{time2} together, 
and simplifying the momenta integral measure to read
\be
\label{intmeasure}
 \int d^3p d^3 p' \delta^3 (\vec{q} + \vec{p} + \vec{p'} ) 
= \frac{2\pi}{q} \int^\infty_0 dp \,p\, \int^{|p+q|}_{|p-q|} dp'\, p',
\ee
we obtain the one-loop correction to be 
\be
\label{oneLoop}
\langle \zeta_{\vec{q}} \zeta_{-\vec{q}} \rangle_{1-\text{loop}} = \frac{H^4 (\tau_q ) }{(2\pi)^2 q^6}
\int^{\infty}_0 dp\, \int^{|p+q|}_{|p-q|} dp'\, \frac{pp'}{q(p+p'+q)} \left(   \frac{1}{q} +  \frac{1}{p+p'+q} \right)
  \mathcal{\tilde{M}} (p,p',q), 
\ee
where 
\bea
&& \textbf{Scalar} : 
\mathcal{\tilde{M}}_s = pp' , \cr \cr
&& \textbf{Fermions} : \mathcal{\tilde{M}}_f = 2 (p-p')^2 (1+ \hat{p} \cdot \hat{p}' ), \cr \cr
&& \textbf{U(1) gauge field}: \mathcal{\tilde{M}}_g = \frac{1}{2}pp' (1+ \hat{p} \cdot \hat{p}' )^2 .
\eea
In \cite{Weinberg1} and \cite{Chai}, these results were stated with the first term in \eqref{oneLoop} absent
due to the omission of the second term in \eqref{oneloopSK}, i.e. $ \int^0_{-\infty_-} d\tau_1 \int^0_{-\infty_+} d\tau_2\,\, \langle H_{\zeta MM} (\tau_1) \zeta^2 (0)  H_{\zeta MM} (\tau_2) \rangle$. As we shall take note shortly,
this term is non-vanishing for the scalar but vanishes for the fermions and gauge fields.

\subsection{Logarithmic terms at one-loop from dimensional and cutoff regularization}


\subsubsection{Logarithmic term from cutoff regularization}
We first consider a quick and blunt way of reading off the logarithmic term by cutoff regularization.
Rescaling all variables by $q$, \eqref{oneLoop} can be written as $\frac{H^4 (\tau_q ) }{(2\pi)^2 q^3} \times I(q)$, with
\be
\label{oneloop}
I(q) =   \int^{L/q}_0 dp\, \int^{p+1}_{|p-1|} dp'\, \left[ F_1(p,p') + F_2 (p,p') \right],
\ee
where $L$ is some UV cutoff, and $F_{1,2}$ correspond to the first and second terms in the bracket of \eqref{oneLoop}. For now, we shall treat it as a `constant' so as to relate our results closely 
to those of \cite{Weinberg1} and \cite{Chai}, yet we bear in mind that it should in principle be $q$-dependent 
and relates to the physical cutoff for dynamics at horizon exit. It is easy to see that the only possible origin of the $\log (q)$ dependence can be traced to the upper integration limit in \eqref{oneloop}. 
Let us begin with the case of the scalar field where
$$
F_1(p,p') = \frac{p^2 p'^2}{p+p'+1},
$$
and the integral evaluated at the upper limit $L/q = \frac{L/\mu}{q/\mu}$ is the only potential origin of any finite logarithmic term.  
Defining $\tilde{L} = \frac{L}{q}$, a straightforward integration yields
\be
\label{I1scalar}
I_1 (\tilde{L}) = \frac{1}{30} \log \tilde{L} +
\frac{1}{30} (1 + \tilde{L} )^3 (1 - 3\tilde{L} + 6 \tilde{L}^2 ) \log \left( 1 + \tilde{L}^{-1} \right) + \ldots ,
\ee
where the ellipses refer to polynomials in $\tilde{L}$. 
In extracting the finite $q$-dependent terms, we can read off from  \eqref{I1scalar}
\be
I_1 = -\frac{1}{30} \log \left( \frac{q}{\mu} \right) + \ldots
\ee
We have adopted a UV cutoff $L$ which isn't manifest in the final expression for the finite logarithm term. This is the term first reported in \cite{Weinberg1} and \cite{Chai}. Similarly, for the 
term corresponding to $F_2 = F_1/(p+p'+1)$, we find
$I_2 (\tilde{L} ) = -\frac{1}{6} \log (1+ \tilde{L} ) + \frac{1}{6}\tilde{L}^3 (4+3\tilde{L} ) \log ( 1+ \tilde{L}^{-1} )+ \ldots$ where we suppress the polynomial terms. Extracting the finite logarithmic term, we obtain
\be
I_2 = \frac{1}{6}  \log \left( \frac{q}{\mu} \right) + \ldots
\ee
Summing the two contributions together, we obtain
\be
\label{oneloopS}
I_{scalar} = I_1 + I_2 = \frac{2}{15}  \log \left( \frac{q}{\mu} \right) + \ldots
\ee
Let us now return to address the UV cutoff $L$ more carefully.
In the above, we have adopted a fixed UV cutoff
and extracted the logarithm term by discarding terms which diverge
as the cutoff is taken to positive infinity. As first pointed out in 
\cite{Senatore},
this momentum cutoff is one
defined for the comoving momentum, and carries the same scaling ambiguity as
that of the scale factor. 
The physical cutoff should be defined in terms of proper
length scales. Denoting $\Lambda_{phy}$
as the `physical' momentum cutoff, we can write
\be
\Lambda_{phy} = \frac{L}{a(\tau_q)},
\ee
where we have used the scale factor at horizon-crossing, noting that the integrals are dominated by dynamics at horizon-crossing where $a(\tau_q) = q/H (\tau_q)$. This implies that we can write the cutoff as
\be
\frac{L}{q}  = \frac{\Lambda_{phy}}{H (\tau_q)}
\ee
leading to the effective replacement in \eqref{oneloopS}
\be
\label{replace}
\log \left( \frac{q}{\mu} \right) \rightarrow \log \left( \frac{H}{\mu} \right),
\ee
where $\mu$ is the renormalizaion scale.\footnote{ Another way to see this as explained in \cite{Senatore} 
is to perform the momentum integral first 
with a time dependent-cutoff as follows. Schematically, for various quantities that one encounters at two-vertices, the cutoff regularization is of the form
\be
I = \int^0_{-\infty_-} d\tau_2\,  \, \int^{\tau_2  e^{\frac{H}{\Lambda}}}_{-\infty_-} d\tau_1  \int^{\Lambda a(\tau_1)} d^3 p
\int d^3 p'  \delta^3 \left(\vec{q} + \vec{p} + \vec{p}' \right) g\left( \vec{p}, \vec{p}', t_1, t_2, \vec{q} \right)
\ee
It was argued in \cite{Senatore} that this yields \eqref{replace}.} 
After taking this subtlety into account, we
eventually have 
\be
I_{scalar} = \frac{2}{15}  \log \left( \frac{H}{\mu} \right) + \ldots
\ee

The computations are similar for the Dirac fermions and $U(1)$ gauge fields, with one notable difference: 
the term that comes from 
the term $ \int^0_{-\infty_-} d\tau_1 \int^0_{-\infty_+} d\tau_2\,\, \langle H_{\zeta MM} (\tau_1) \zeta^2 (0)  H_{\zeta MM} (\tau_2) \rangle$ in \eqref{oneloopSK} turns out not to contain any finite logarithmic term. 
Explicitly, suppressing the unimportant polynomial terms, for the fermions, we have 
\bea
I_1 (\tilde{L} ) &=& \frac{2}{15} \log (1 + \tilde{L} ) + 
\left( 2\tilde{L}^2 + \frac{20}{3}\tilde{L}^3 + 8\tilde{L}^4 + \frac{16}{5} \tilde{L}^5 \right ) \log ( 1 + \tilde{L}^{-1} ) + \ldots\cr
I_2 (\tilde{L}) &=& \left(
2\tilde{L} + 10 \tilde{L}^2 + 16 \tilde{L}^3 + 8 \tilde{L}^4
\right) \log ( 1+ \tilde{L}^{-1} ) + \ldots.
\eea 
whereas for the U(1) gauge field, we find 
\bea
I_1 (\tilde{L} ) &=& \frac{1}{15} \log (1 + \tilde{L} ) + \left(  \frac{2}{3}\tilde{L}^3 + \tilde{L}^4 + \frac{2}{5}\tilde{L}^5  \right) \log (1 + \tilde{L}^{-1} )  + \ldots \cr
I_2 (\tilde{L} ) &=&  \left( \tilde{L}^2 + 2\tilde{L}^3
+ \tilde{L}^4 \right) \log (1 + \tilde{L}^{-1} ) + \ldots .
\eea 
Below, we summarize 
the various results
\bea
\label{Oscalar}
I_{scalar} &=& \frac{2}{15} \log \left( \frac{H}{\mu} \right)  + \ldots, \\
\label{Ofermion}
I_{fermion} &=& -\frac{2}{15}  \log \left( \frac{H}{\mu} \right)  + \ldots, \\
\label{Ogauge}
I_{gauge} &=& -\frac{1}{15}  \log \left( \frac{H}{\mu} \right)  + \ldots
\eea
Thus we see that the one-loop contribution of the Dirac fermion exactly cancels that of the scalar 
whereas the correction due to the $U(1)$ gauge field is half that of the Dirac fermion and of the same (negative) sign.
Before we consider \eqref{Oscalar} - \eqref{Ogauge} using dimensional regularization, we should remark that \eqref{Oscalar} differs from 
that reported in \cite{Senatore} and \cite{Eugene} which presumably corrected the one-loop term derived in \cite{Weinberg1,Chai} by 
including $I_2$ and (in \cite{Senatore} ) by taking into account dimensional regularization of the scale factor and wavefunctions. 
The $I_1$ term we computed here is identical to that stated in \cite{Weinberg1,Chai} modulo the logarithmic argument. The one-loop terms for fermions and gauge fields were presented in \cite{Chai} with the $I_2$ term not being taken into account at all even though they turn out to vanish. While \eqref{Ofermion} is identical to that presented in \cite{Chai}, \eqref{Ogauge} differs. In the following, we check our results by deriving them using a different regularization method.

\subsubsection{Logarithmic one-loop corrections from dimensional regularization }  

We now study if we recover the same finite logarithmic term in dimensional regularization
as a consistency check. We begin by 
writing the spatial dimensionality as
$
d = 3+ \delta,
$
and noting that the angular integration should generalize as 
\be
\label{angular}
\int d\Omega_{2+\delta} = \int^\pi_0 d\Theta\, \, \sin^{1+\delta} \Theta \times \text{Vol} (S_{1+\delta} ) = 
\int^\pi_0 d\Theta \sin^{1+\delta} \Theta \times  \frac{(2\pi)\pi^{\delta/2}}{\Gamma \left(1 + \frac{\delta}{2}\right)}. 
\ee
Keeping aside the dimensional regularization factors associated with the scale factors and mode wavefunctions for the moment,
we then have 
\bea
I (q) &=&   \frac{1}{2\pi q^3}    \int d^{3+\delta}p \int d^{3+\delta} p' \,\, \delta^{3+\delta} ( p+ p' + q) \left[ F_1 (p,p') + F_2 (p,p') \right] \cr
&=&
\label{comDim}
\frac{q^{\delta}}{2\pi}  
\int^{2\pi}_0 d\phi\,\, \sin^\delta \phi\,\,
\int^{\infty}_0 dp\,\, p^{\delta} \, \int^{p+1}_{|p-1|} dp'\,\, \sin^\delta \Theta 
\left[ F_1 (p,p') + F_2 (p,p') \right] \cr
&=&
\label{comDim}
q^\delta  \left(   \frac{\pi^{\delta /2}}{\Gamma (1 + \delta /2)}    \right)
\int^{\infty}_0 dp\,\, p^{\delta} \, \int^{p+1}_{|p-1|} dp'\,\, \sin^\delta \Theta 
\left[ F_1 (p,p') + F_2 (p,p') \right],
\eea
with 
$
\sin^\delta \Theta = \left[  1 - \left( \frac{(p')^2 - p^2 -1}{2p}     \right)^2   \right]^{\delta/2}
$.
To proceed, we find it useful to perform a coordinate transformation which is singular at zero $\delta$ as follows. 
\be
P = p^\delta, \qquad dp = \frac{p}{\delta} \frac{dP}{P}.
\ee
Let $J_{1,2} (\delta )$ denote
the momenta integrals in \eqref{comDim} corresponding to the integrand terms $F_{1,2}(p,p')$ respectively, 
with $\sin^\delta \Theta$ term absent and in the domain $p>1$. 
In the case of the scalar field,
we find
\bea
J_1 (\delta ) &=& \int^{\infty}_1 dp\,\, p^\delta p^2 \left(     
-2 - (p+1)^2 \log \frac{p}{p+1}
\right) \cr
\label{logS}
&=&  
\int^{\infty}_1 dP\,\, \frac{P^{3/\delta}}{\delta} \left( -2 + (P^{1/\delta} +1)^2 \log \left(   
1 + P^{-1/\delta}
\right) \right),
\eea
where in the limit $\delta \rightarrow 0$, we find this integral yields a cut-off independent finite term 
that reads
\be
\label{polesingular1}
J_1 (\delta ) = -\frac{1}{30\delta} + \ldots 
\ee
which is identical to the result obtained in the previous section via cutoff regularization. 
One can locate this term from the few first terms in the Taylor expansion of the logarithm term in \eqref{logS}, 
considering only those that give a $\delta$-independent terms in the integrand. 
Substituting \eqref{polesingular1} into equation \eqref{comDim} precisely reproduces the logarithmic term we obtained previously using cutoff regularization.
The other $\delta$-dependent terms in \eqref{comDim}, apart from $q^\delta$,  do not contribute to this logarithmic running apart from unimportant numerical constants. 
Explicitly, the factor 
$$
\frac{\pi^{\delta /2}}{\Gamma ( 1 + \frac{\delta}{2} ) } \approx 1 +\delta \left( \frac{\gamma}{2} + \frac{1}{2} \text{Log} ( \pi ) \right) + \ldots,
$$
whereas expanding the $\sin^\delta \Theta$ term introduces a term 
$
\text{Log}  \left( 1 - \frac{(r^2-p^2-1)^2}{4p^2} \right)
$
in the integrand, which gives rise to a numerical constant. Similarly for the fermions and vector fields, these terms do not play 
a role in yielding nontrivial one-loop correction terms, and henceforth, we omit them in our discussion. 

Now for the term corresponding to $F_2 (p,p')$, we obtain 
\bea
J_2 (\delta ) &=& \int^{\infty}_1 dp\,\, \frac{1}{2} p^{\delta +1} 
 \left(     
1 + 5p + 4p (1+p)
 \log \frac{p}{p+1}
\right) \cr
&=&  \int^{\infty}_1 dP\,\, \frac{P^{2/\delta}}{2 \delta} \left( 1 + 5 P^{1/\delta}  + 4
P^{1/\delta} \left( 1 + P^{1/\delta} \right)
 \log \left(   
1 + P^{-1/\delta}
\right) \right), 
\eea
where in the limit $\delta \rightarrow 0$, we find this integral diverges but it contains a finite term  
that reads
\be
\label{polesingular}
J_2 (\delta ) = \frac{1}{6\delta} + \ldots 
\ee
which is again identical to the result obtained via cutoff regularization. Naively then, we have the one-loop term 
being $I (q) = \frac{2}{15} \log \left( \frac{q}{\mu} \right)$, after expanding $q^\delta = 1 + \delta \, \log (q) + \ldots$.
 
Finally, we take into account the
analytic continuation of the scale factors in the integration measure and the wavefunctions ( those running in the loop and not external ones). This was already explained in \cite{Senatore} and thus let us very briefly elaborate on this point with some details that are relevant specifically for our context. For each vertex, the analytic continuation of the interaction Hamiltonian introduces a factor of $a^\delta (\tau )$ due to the dependence on the metric determinant, each of which can be expanded as 
\be
a^\delta (\tau ) = 1 - \delta \log ( -H \tau ) + \ldots
\ee
The time-integrals are dominated by the time at horizon exit, and thus for each scale factor, this modifies our final result for the one-loop logarithmic term to be
\be
I(q) = q^{\delta} \left(  \frac{2}{15\delta } + \ldots \right) \times \left( 1 - \delta \log \left(\frac{H}{q} \right) + \ldots \right).
\ee
Apart from this correction, the analytic continuation of the scalar wavefunctions running in the loop yields for each an additional factor of $\frac{1}{2} \delta \log ( - H \tau )$ coming from the fact that 
\be
\zeta (\tau ) \sim \frac{H^{1+ \delta /2}  (-k \tau )^{(3+\delta )/2}}{k^{(3+\delta )/2}} H^{(1)}_{(3+\delta )/2} (-k\tau )
\ee
and similarly for $\chi (\tau )$. Since there are six such factors, summing all corrections, one obtains
\be
I(q) = \left( 1 + \delta \log \left( \frac{q}{\mu} \right) + \ldots \right) \left(  \frac{2}{15\delta } + \ldots \right) \times \left( 1 +\left( \frac{6}{2} -2 \right) \delta \log \left(\frac{H}{q} \right) + \ldots \right) = \frac{2}{15} \log \left(  \frac{H}{\mu} \right) + \ldots
\ee
which is identical to the result obtained via cutoff regularization.

Let us complete our discussion by analyzing the fermions and gauge fields' cases: we find that in both cases,
the term $J_2 (\delta )$ does not yield any finite logarithmic term, whereas the $J_1 (\delta )$ term is non-vanishing and reads in each case:
\begin{enumerate}
\item \textbf{Fermions}: 
\bea 
J_1 (\delta ) &=& \int^{\infty}_1 dp\,\, \frac{2}{3} p^{\delta } 
 (1+2p)
\left(     
1 + 12p + 12p^2 
+6p (1+3p+2p^2 )
 \log \frac{p}{p+1}
\right) \cr
&=&  \int^{\infty}_1 dP\,\, \frac{2P^{\frac{1}{\delta}}}{3 \delta} 
 (1+2P^{\frac{1}{\delta}})
\left(     
1 + 12P^{\frac{1}{\delta}} + 12P^{2/\delta} 
+6 P^{\frac{1}{\delta}} (1+3P^{\frac{1}{\delta}}+2P^{2/\delta} )
 \log \left(1+ P^{-\frac{1}{\delta}}  \right)
\right).\nonumber\\
\eea
\item \textbf{U(1) gauge fields}:
\bea
J_1 (\delta ) &=& \int^{\infty}_1 dp\,\,p^\delta
 \left[     
\frac{1}{6} - \frac{2p}{3} - 3p^2 -2p^3 -
2p^2 (1+p)^2 \log \frac{p}{1+p}
\right] 
\cr
&=&  \int^{\infty}_1 dP\,\, \frac{P^{\frac{1}{\delta}}}{\delta} 
\left[     
\frac{1}{6} - \frac{2 P^{\frac{1}{\delta}} }{3} - 3P^{2/\delta} -2P^{3/\delta} -
2P^{2/\delta} (1+P^{\frac{1}{\delta}})^2 \log \left( 1+P^{-\frac{1}{\delta}} \right)
\right]. \nonumber \\
\eea
\end{enumerate}
They give rise to the singular terms $-\frac{2}{15\delta}$ and $-\frac{1}{15\delta}$ respectively which precisely agree with the previous calculation based on cutoff regularization. 

Similar to the scalar field, one needs to analytically continue the wavefunctions in dimensional regularization. For the fermions, as pointed out in for example \cite{Chen}, in $d+1$-dimensional de Sitter, the Dirac equation turns out to read $i \left(\gamma^\mu \partial_\mu - \frac{d}{2\tau} \gamma^0 \right)\Psi = 0$ which admits the $d-$dimensional spinor wavefunction 
\be
\Psi (\vec{x}, t) = (-H\tau )^{\frac{d}{2}}  \int d^dq\,\, \sum_s \, e^{i\vec{q} \cdot \vec{x}} 
\left[   U_{\vec{q}, s} (t) a_{\vec{q},s} + V_{-\vec{q}, s} (t) \beta^\dagger_{-\vec{q}, s}
\right], 
\ee
where $U_{\vec{q},s }, V_{\vec{q},s}$ are spinors with definite conformal momenta in $d+1$-dimensional Minkowski spacetime. Expanding 
$d=3+\delta$, one obtains a correction factor of $\frac{1}{2} \delta \log (-H\tau) $ for the logarithmic running in $I(q)$, similar to what arises
from the analytic continuation of scalar wavefunctions. For the gauge fields in the parametrization of \eqref{m3}, we checked that the free
$d-$dimensional Maxwell equations in conformal coordinates imply that the modes $A_q (\tau )$ satisfy 
\be
\frac{d^2A_q}{d\tau^2} + q^2 A_q - (d-3)\frac{1}{\tau} \frac{dA_q}{d\tau} = 0.
\ee
In the case of $d=3$, we obtain plane waves which can be normalized with Bunch-Davies condition. 
For generic $d= 3+\delta$,
we find the solution  
\be
A_q \sim (-H\tau )^{\frac{1+\delta}{2}} H^{(1)}_{\frac{1+\delta}{2}} (-q\tau),
\ee
where $H^{(1)}_v (-q\tau )$ denotes the Hankel function of the first kind. 
Once again expanding $d = 3+\delta$, one obtains the same correction factor of $\frac{1}{2} \delta \log (-H\tau) $ for the logarithmic running in $I(q)$. Together with our computations of $J_{1,2} (\delta )$,
this implies that for the fermions and gauge fields, the one-loop logarithmic running goes as 
$$
I_{fermion} = -\frac{2}{15} \log \left(  \frac{H}{\mu}  \right), \qquad
I_{gauge} = - \frac{1}{15} \log \left(  \frac{H}{\mu}  \right)
$$
which are identical to the results obtained via cutoff regularization. Finally, we recall that
the tree-level primordial power spectrum is 
$
\langle \zeta \zeta \rangle_{tree} = \frac{H^2 / M^2_{pl}}{2(2\pi)^3 q^3 \epsilon},
$
and thus all the various one-loop correction terms $\langle \zeta \zeta \rangle_{1-\text{loop}} = \frac{H^4 (\tau_q )}{(2\pi)^2 q^3} I (q) $
are suppressed by $\sim \epsilon \frac{H^2}{M^2_{pl}}$.


\section{On noncommutative QFT and inflationary dynamics}

\subsection{Some aspects of noncommutative quantum field theory}

In the following, we very briefly review a couple of aspects of noncommutative QFT (see for example \cite{Minwalla} for a more extensive discussion). 
We begin with the commutation relation for a noncommutative $\mathbb{R}^d$ 
\be
[ x^\mu, x^\nu ] = i \theta^{\mu \nu}
\ee
where $\theta^{\mu \nu}$ is a real, constant and antisymmetric matrix. 
In previous literature notably \cite{Ho,Cai,Akofor1,Nautiyal}, $\theta^{0i}$ was turned on and shown to
deform the two-point function at tree-level.  For simplicity, we will not consider a non-vanishing $\theta^{0i}$ 
in this work, only switching on the spatial noncommutativity (see \cite{Palma} for a related discussion in the context
of the one-loop effective action for inflation on a noncommutative spacetime).\footnote{We note in passing that in \cite{Gomis,Ganor}, it was
claimed that a non-zero $\theta^{0i}$ may pose problem with unitarity at the field theoretic level and that there 
appears to be no regime of parameters where a string theoretic construction could be associated with it essentially 
because it necessarily brings with it massive string states.} 

The algebra of functions on noncommutative $\mathbb{R}^d$ is an algebra 
of ordinary functions on $\mathbb{R}^d$ with the product deformed to a noncommutative but associative
Moyal star product defined by 
\be
\left( \phi_1 \star \phi_2 \right) (x) = e^{\frac{i}{2} \theta^{\mu \nu} \partial^y_\mu \partial^z_\nu } \phi_1 (y) \phi_2 (z) \vert_{y=z=x} .
\ee
A practical formulation of noncommutative QFTs is to adopt as the starting point an action of the same form but with the fields in the Lagrangian being multiplied using the above Moyal star product. The quadratic part of the action remains the same as in the commutative theory since 
\be
\int d^dx \, \partial \phi (x) \star \partial \phi (x) = \int d^d x \, \partial \phi (x) \, \partial \phi (x) ,
\ee
where total derivative terms have been dropped using suitable boundary conditions on $\phi$. Propagators thus take the identical form as in the commutative theories. On the other hand, interactions will be generically modified.
For example, noncommutative polynomial interactions take the form 
\be
\label{productMoyal}
\phi_1 (x_1) \star \ldots \star \phi_n (x_n) = \prod_{a<b} \text{exp} \left( \frac{i}{2} \theta^{ \mu \nu} \frac{\partial}{\partial x^\mu_a} \frac{\partial}{\partial x^\nu_b} \right)
\phi_1 (x_1) \ldots \phi_n (x_n).
\ee
In particular it is useful to note the cyclic permutation symmetry of Moyal product inside the integral, i.e.
$\int d^4x \, f(x) \star h(x) \star g(x) = \int d^4x\, h(x) \star g(x) \star f(x)$. Thus, for example, the
noncommutative deformation of the complex scalar field theory with potential $ V(\phi, \bar{\phi} )= ( \bar{\phi} \phi )^2$ 
is described by a two-coupling deformation $V (\phi, \bar{\phi} ) = g_1 \phi \star \bar{\phi} \star \phi \star \bar{\phi} + 
g_2 \phi \star \phi \star \bar{\phi} \star \bar{\phi} $. In noncommutative QED, the gauge-fermion coupling is generalized to the form
$
\overline{\varphi} \star \gamma^\mu A_\mu \star \varphi - \overline{\varphi} \star \varphi \star \gamma^\mu A_\mu
$
which is compatible with the noncommutative gauge transformation of the form $\delta_\lambda A_\mu = \partial_\mu \lambda - ie
(A_\mu \star \lambda - \lambda \star A_\mu )$. 

\subsection{Noncommutative deformations of the interaction Hamiltonians}

In the following, we construct noncommutative deformations of the interactions between minimally coupled matter fields and the scalar fluctuation of the metric $\zeta$. As we have seen, the various interaction Hamiltonians ( \eqref{H1}, \eqref{H2}, \eqref{H3} ) are trilinear in the fields and 
can be read off from \eqref{Hint} as derived in \cite{Weinberg1,Chai}. 
We begin by considering a two-coupling ($g,h$) noncommutative deformations of the following form  
\be
H_{int} \sim g M (x) \star N (x) \star \nabla^{-2} \dot{\zeta} + h M (x) \star \nabla^{-2} \dot{\zeta} \star N (x) ,
\ee
where $M(x), N(x)$ represent matter fields with appropriate derivative operators acting on them. In the limit of vanishing 
noncommutative parameter, we wish to recover the usual interaction Hamiltonian, and hence we impose the moduli constraint
\be
\label{constraint1}
g + h = 1.
\ee
Another constraint is more visible when we work in momentum space. Schematically writing 
$
M(x) = \int d^3p \,\,M_{\vec{p}} (t) \, e^{i \vec{p} \cdot \vec{x}}
$
where $M_{\vec{p}}$ is the field in momentum space, the interaction Hamiltonian then reads 
\be
H_{int} \sim \int d^3p \int d^3p' \delta^3 (\vec{p} + \vec{p'} + \vec{q} ) \left( g e^{p' \wedge p} + h e^{-p' \wedge p} \right) M_{\vec{p'}} (t) 
N_{\vec{p}} (t) \left( \nabla^{-2} \dot{\zeta}\right)_{\vec{q}} (t),
\ee
in which we denote $
p' \wedge p \equiv -\frac{i}{2} \theta_{m n} p^n {p'}^m 
$. Since $H_{int}$ should be real, we obtain 
\be
\label{constraint2}
\bar{g} = h.
\ee
Together with \eqref{constraint1}, we are led to a one-parameter ($g$) noncommutative deformation with 
\be
\label{constraint3}
\text{Re} (g) = \frac{1}{2}.
\ee 
We note in passing that other models of noncommutative deformations of inflationary dynamics have appeared in past literature. In \cite{Mota}, noncommutative spacetime geometry is understood in the form of a deformation of the canonical commutation relation
between the field operators and this could lead to phenomenological consequences different from what we obtain from our construction of noncommutative deformation. For example, it turns out that a non-vanishing $\theta^{0i}$ leads to the primordial power spectrum attaining a factor of $e^{H \theta^{0m} k^m}$ as discussed in \cite{Mota,Calmet} instead of $\cosh (H \theta^{0m} k^m)$ as in \cite{Akofor1}.


\subsection{Planar and non-planar graphs at one-loop}

We now revisit the one-loop contributions to $\langle \zeta \zeta \rangle$ due to the noncommutative trilinear interactions. For definiteness,
let's begin with the term $\langle H_{int} (\tau_1 ) H_{int} (\tau_2 ) \zeta \zeta \rangle$  - an eight-point function which factorizes into a product of two four-point functions, one involving the external matter fields and the other one involving the $\zeta$'s. Each Wick contraction yields a specific Moyal phase factor, 
with non-trivial ones that can be interpreted diagrammatically as non-planar graphs and trivial ones being planar ones. 
We find that fermions and complex scalars attain noncommutative corrections that can be described purely by 
non-planar diagrams whereas real scalars and vector fields always have corrections that are described by a sum of planar and non-planar
diagrams. 

Below, we derive this point in detail by manifestly keeping track of the Wick contractions among the matter fields in the four-point function and using \eqref{productMoyal}. The overall phase factor depends on the relative position of each field in the correlation function and its contraction with the other fields. To simplify notations, we omit all derivative operators acting on the fields which do not affect the computation of Moyal phase factors.  

\begin{figure}[h]
\centering
\includegraphics[width=120mm]{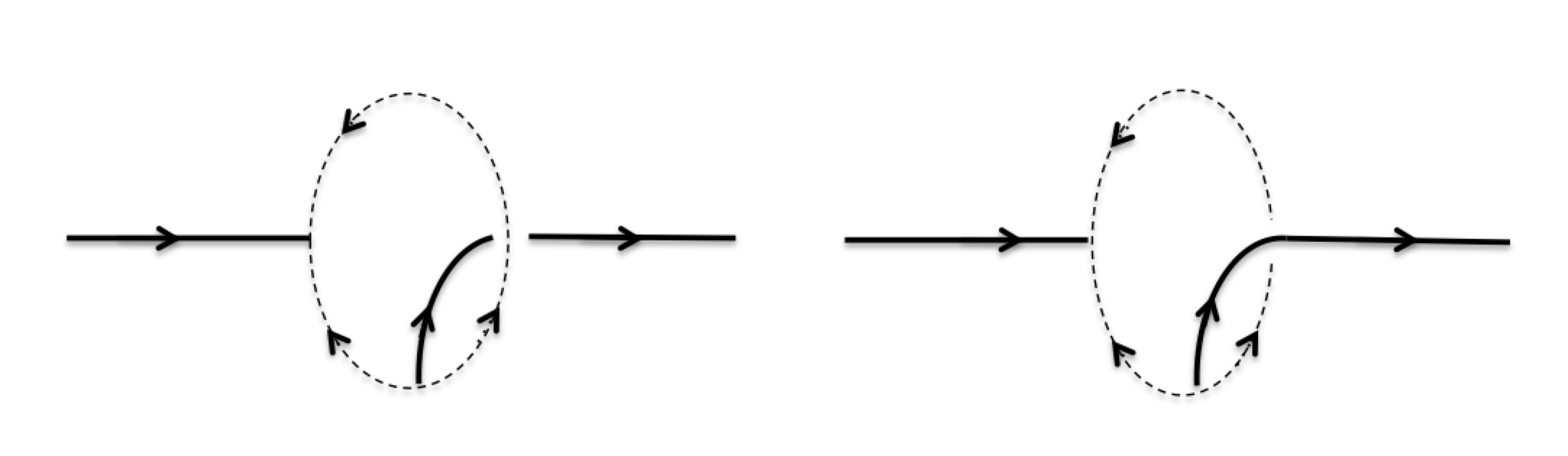}
\vspace{10pt}
\put(-25,35){$\zeta_{\vec{q}}$}
\put(-155,35){$\zeta_{\vec{q}}$}
\put(-205,35){$\zeta_{\vec{q}}$}
\put(-325,35){$\zeta_{\vec{q}}$}
\put(-255,85){$X_{\vec{p'}}$}
\put(-285,15){$X_{\vec{p}}$}
\put(-85,85){$X_{\vec{p'}}$}
\put(-118,15){$X_{\vec{p}}$}
\caption{We denote $X_{\vec{P}}$ to represent a massless matter field $X$ with momentum $\vec{P}$. The non-planar diagram on the left 
corresponds to terms which attain a phase factor of $2p' \wedge p$ whereas the one on the right is associated with the conjugate phase factor
$-2p' \wedge p$. Both planar and non-planar diagrams are present in the one-loop correction. For the real scalar and abelian gauge field, they
yield an overall deformation factor $\frac{1}{2} ( 1 + \cosh (2p' \wedge p)$ to the commutative one-loop correction. For the massless fermion, 
there is a choice of noncommutative deformation such that the overall factor is simply $ \cosh (2p' \wedge p)$, i.e. only non-planar diagrams contribute.}
\label{fig2}
\end{figure}
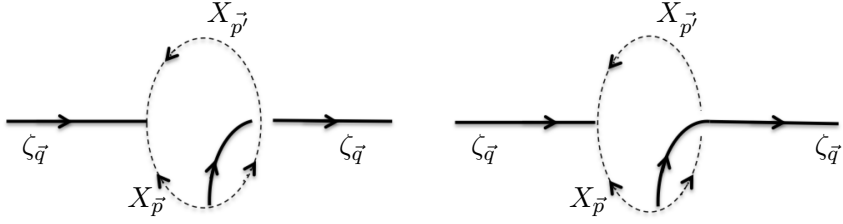

First, consider the two cases (fermions and complex scalars) for which we could only have non-planar diagrams. For fermions, we have $M(x) = \overline{\Psi} (x) $ and $N(x) = \Psi (x)$ whereas for the complex scalar, we have $M(x) = \bar{\Phi } (x), N(x) = \Phi (x)$.  For each of the four terms in $\langle H_{int} (\tau_1 ) H_{int} (\tau_2 ) \zeta  \zeta \rangle$, denoting the internal momenta of $M(x_1), N(x_1)$ to be $p', p$ respectively,
we find the phase factors to read
\bea
\text{Moyal phase of\,\,\,\,} \bcontraction[2ex]{}{M} {(x_1) \star N (x_1) \star \zeta (x_1) \star
M (x_2) \star }{N} 
\bcontraction{M (x_1) \star}{ N}{ (x_1) \star \zeta (x_1) \star}{ M}  
M(x_1) \star N(x_1) \star \zeta (x_1) \star M(x_2) \star N(x_2) \star \zeta (x_2)
\qquad &\sim& 0 \cr \cr
\bcontraction[2ex]{}{M} {(x_1) \star N (x_1) \star  \zeta (x_1) \star M (x_2) \star \zeta (x_2)}{N} 
\bcontraction{M (x_1) \star }{ N}{ (x_1) \star  \zeta (x_1) \star }{ M}  
M(x_1) \star  N(x_1)  \star  \zeta (x_1) \star  M(x_2)  \star  \zeta (x_2)  \star  N(x_2)
\qquad &\sim& 2 p' \wedge p \cr \cr
\bcontraction[2ex]{}{M} {(x_1)  \star  \zeta (x_1) \star  N (x_1) \star M (x_2) \star  }{N} 
\bcontraction{M (x_1) \star  \zeta (x_1) \star   }{ N}{ (x_1) \star }{ M}  
M(x_1)   \star  \zeta (x_1) \star   N(x_1)   \star   M(x_2) \star   N(x_2)  \star  \zeta (x_2)
\qquad &\sim& -2 p' \wedge p \cr\cr
\label{table1}
\bcontraction[2ex]{}{M} {(x_1)   \star   \zeta (x_1)  \star  N (x_1)  \star  M (x_2)  \star   \zeta (x_2) \star }{N} 
\bcontraction{M (x_1) \star   \zeta (x_1)  \star  }{ N}{ (x_1) \star }{ M}  
M(x_1) \star  \zeta (x_1)  \star  N(x_1) \star  M(x_2) \star  \zeta (x_2)  \star  N(x_2)
\qquad &\sim& 0.
\eea
The Wick contractions among the $\zeta$'s do not change the phase factors and hence we omit them above. 
This implies then that 
the noncommutative
one-loop term  $\langle H_{int} (\tau_1 ) H_{int} (\tau_2 ) \zeta \zeta \rangle$ gains the following correction factor in momentum space 
\be
\label{oneloopM1}
 \langle H_{int} (\tau_1 ) H_{int} (\tau_2 ) \zeta  \zeta \rangle_{(p',p)}  \rightarrow  \left[  2|g|^2  \cosh (2p' \wedge p ) + g^2 + \bar{g}^2 \right] \langle H_{int} (\tau_1 ) H_{int} (\tau_2 ) \zeta  \zeta \rangle_{(p',p)}.
\ee
The $\theta$-independent term in the square bracket of 
\eqref{oneloopM1} is depicted by the planar diagram equivalent 
to the commutative case whereas the other term can be represented by non-planar diagrams 
that we will study in detail in the following section. 
Also, the two points in the modulus space corresponding to the absence
of planar diagrams are 
\be
g_{\text{non-planar}} = \frac{1}{\sqrt{2}}e^{\pm \frac{i\pi}{4}}.
\ee
They can be mapped to each other by switching the sign of $\theta$.  
For the real scalar field, due to cyclic permutation symmetry, the noncommutative Hamiltonian reads
\be
H_{\zeta \chi \chi} = -\int d^3x\, \epsilon H a^5 \left(  2 \dot{\chi} \star \dot{\chi} \star \nabla^{-2} \dot{\zeta} \right). 
\ee
Keeping track of the Moyal phase factor for each fully contracted term, we find 
\bea
\acontraction[2ex]{}{\chi} {(x_1) \star \chi (x_1) \star  \zeta (x_1) \star }{\chi} 
\bcontraction[2ex]{ \chi (x_1) \star }{ \chi }{ (x_1) \star \zeta (x_1) \star \chi (x_2) \star}{ \chi }  
\chi (x_1)  \star  \chi (x_1) \star  \zeta (x_1) \star \chi (x_2) \star \chi (x_2) \star \zeta (x_2)
\qquad &\sim& 2 p' \wedge p, \cr \cr
\acontraction[2ex]{}{\chi} {(x_1) \star \chi (x_1) \star  \zeta (x_1) \star \chi (x_2)   \star      }{\chi} 
\bcontraction[2ex]{ \chi (x_1) \star }{ \chi }{ (x_1) \star \zeta (x_1) \star }{ \chi }  
\chi (x_1)  \star  \chi (x_1) \star  \zeta (x_1) \star \chi (x_2) \star \chi (x_2) \star \zeta (x_2)
\qquad &\sim& 0.
\eea
After taking into account the fact that the rest of the correlation function in momentum space is symmetric w.r.t. $p' \leftrightarrow p$,
we find that each term in \eqref{table1} gains an identical correction factor $1 + \cosh (2p' \wedge p )$, and thus 
\be
\label{oneloopM2}
 \langle H_{int} (\tau_1 ) H_{int} (\tau_2 ) \zeta  \zeta \rangle_{(p',p)}  \rightarrow  \frac{1}{2} \left[ 1 + \cosh (2p' \wedge p )  \right] \langle H_{int} (\tau_1 ) H_{int} (\tau_2 ) \zeta  \zeta \rangle_{(p',p)}.
\ee
The $\theta$-independent part of \eqref{oneloopM2} can be represented by planar diagram whereas the part containing 
$ \cosh (2p' \wedge p )$ is depicted by 
the non-planar diagrams. 
In the commutative case, the complex scalar field is equivalent to two real scalars via a field redefinition
but this equivalence is however spoilt by the noncommutative deformation. Also, we find that \eqref{oneloopM2} also 
applies to the $U(1)$ gauge field for which the noncommutative interaction Hamiltonian reads
\be
\label{H3}
H_{\zeta A A} = -\int d^3 x\,\epsilon H a^5 \left(
\frac{1}{a^2}\dot{A}_i \star \dot{A} \star \nabla^{-2} \dot{\zeta}
+ \frac{1}{2a^4} F_{ij} \star F_{ij} \star \nabla^{-2} \dot{\zeta}
\right)  
\ee
We now proceed to compute the one-loop corrections with the 
Moyal phase factors, i.e. the non-planar diagrams as drawn in Figure \ref{fig2}. 

\subsection{The generic Moyal deformation factor for cubic graph contribution to one-loop correction}

We now consider the effect of the noncommutative deformation factor $\cosh (2p' \wedge p)$ and evaluate 
the non-planar diagrams in Figure \ref{fig2} explicitly. 
In the usual commutative setting, 
denoting $\vec{q}$ to be the external momentum (of $\zeta$) and $\vec{p}', \vec{p}$ to be the internal momenta running in the loop,  
one finds that the internal momenta integrals can be conveniently performed via \eqref{intmeasure}
where we took $\vec{q}$ to lie along the $p_z$-axis and noted that
the polar angular integration simply yields a factor of $2\pi$ since the correlation function depends on the magnitude of various momenta.
For the non-planar diagrams in the noncommutative setting, the Moyal phase factor introduces a dependence on the polar angle, breaking the rotational symmetry
so \eqref{intmeasure} is no longer valid generally.  
Since $\vec{q} + \vec{p} + \vec{p}' =0$, we have
\be
p' \wedge p = -\frac{i}{2} \theta_{m n} p^n {p'}^m = \frac{i}{2} \theta_{mn}p^n q^m,
\ee
and further taking $\vec{q}$ to lie along the $p_z$-axis, 
the Moyal phase factor is then 
$$
\cos \left( q \bigg(\theta_{31} (p \sin \Theta \cos \varphi) + \theta_{32} (p \sin \Theta \sin \varphi) \bigg) \right).
$$
Revisiting the momenta integrals in \eqref{intmeasure}, we first perform the polar angle integration which yields
\be
\int^{2\pi}_0 d\varphi \,\, \cos \left( q \bigg(\theta_{31} (p \sin \Theta \cos \varphi) + \theta_{32} (p \sin \Theta \sin \varphi) \bigg) \right) =
2\pi J_0 \left( pq |\theta | |\sin \Theta| \right),
\ee
where 
$$
\theta \equiv \sqrt{\theta^2_{31} + \theta^2_{32}},\qquad |\sin \Theta| = \sqrt{ 1 - \frac{(p'^2 - p^2 - q^2)^2}{4p^2 q^2}  },
$$
and we have invoked a useful integral representation of the Bessel function
\be
\int^{2\pi}_0 d\varphi \cos (a \cos \varphi + b \sin\varphi )  = 2\pi J_0 ( \sqrt{a^2+b^2} ).
\ee
This implies that instead of \eqref{oneloop}, the one-loop correction arising from non-planar graphs
reads
\be
\label{oneloop2}
I(q,\omega) =   \int^{L/q}_0 dp\, \int^{p+1}_{|p-1|} dp'\, \left[ F_1(p,p') + F_2 (p,p') \right] J_0 (p\, \omega \sin \Theta ),
\ee
where $\omega$ is the noncommutative parameter in units of $q$, i.e. $\omega \equiv q^2 \theta$. 
Defining $\theta^i \equiv \frac{1}{2} \epsilon^{ijk}\theta_{jk}$, 
we see that at least
at one-loop order, the noncommutative 
deformation picks up only the magnitude of the noncommutative vector $\theta^i$ projected onto the plane transverse to momentum $\vec{q}$. The form of $I(q,\omega)$  in \eqref{oneloop2} implies that it is a function of $L/q$ and $\omega$. Restoring the generic direction of $\vec{q}$,
the transverse noncommutativity parameter reads 
$$
\theta^{T} = \theta \sqrt{1 - (\hat{q} \cdot \hat{\theta} )^2}, \qquad \theta \equiv |\vec{\theta}|.
$$
For the rest of the paper, we omit the superscript on $\theta$ for notational simplicity. 
We see that the Moyal phase factor implies that the one-loop correction breaks spatial isotropy and 
could be a source of some degree of 
anisotropy of the CMB after convoluting with appropriate transfer functions for which we expect to be separated in scale and hence unmodified by spacetime noncommutativity.\footnote{
Explicitly, expanding the temperature
deviations in terms of spherical harmonics
$
\frac{\delta T (\hat{q})}{T} = \sum_{l,m} a_{lm} Y_{lm} (\hat{q}), 
$
the two-point function $\langle \zeta_{\vec{q}} \zeta_{\vec{q}'} \rangle$ is related to observables via 
$$\langle a_{lm} a_{l' m'} \rangle =  16 \pi^2 (-i)^{l-l'} 
\int \int \frac{d^3q d^3q'}{(2\pi)^6} \Delta_l (q) \Delta_{l'} (q') \langle \zeta_{\vec{q}} \zeta_{\vec{q}'} \rangle {Y^*}_{lm}(\hat{q}) Y_{l' m'}(\hat{q}')
,$$ 
where $\Delta_{l} (q)$ are the transfer functions. See for example \cite{Akofor1}.}

\section{One-loop correction from non-planar graphs}

We now come to the central part of this paper where we study the one-loop correction in \eqref{oneloop2}. 
To make further computations tractable, it is useful to invoke asymptotic expansion of the Bessel function 
for both small and large arguments. For small argument $z<1$, 
$J_0 (z)$ goes as \cite{DLMF}
\be
\label{BesselExpandsmall}
J_0 (z) = \sum^\infty_{k=0} (-1)^k \frac{(\frac{1}{4}z^2)^k}{k! k!},
\ee
whereas for large argument $z \equiv 1/l$, $l<1$, we have \cite{DLMF}
\be
\label{BesselExpandlarge}
J_0 (1/l) = \sqrt{\frac{2l}{\pi}} \left[  \cos \left(  \frac{1}{l} - \frac{\pi}{4} \right) \sum_{k=0}^\infty (-1)^k a_{2k} l^{2k} - 
\sin \left( \frac{1}{l} - \frac{\pi}{4} \right) \sum_{k=0}^\infty (-1)^k a_{2k+1} l^{2k+1} \right],
\ee
with the coefficients being 
$$
a_k  = \frac{(2k-1)!! (2k-1)!! (-1)^k}{k! 8^k}.
$$
From \eqref{oneloop2}, we see that in the limit $\omega = 0$, the Bessel function smoothly goes to unity
and we should recover the one-loop corrections terms in the commutative setting. In the following, we consider the cases
 of $0< \omega < \frac{q}{L}$ and $1> \omega \geq \frac{q}{L}$ separately.  For the former, we can invoke 
\eqref{BesselExpandsmall} throughout the (two-dimensional) domain of integration, whereas for the latter, 
 it is convenient to split the integral into a few pieces where in each of them, either \eqref{BesselExpandsmall} or \eqref{BesselExpandlarge} 
can be used separately depending on whether the argument
$
p \omega \sin \Theta  
$
is smaller or larger than unity. Scaling all virtual momenta to be units of $q$,
using 
$
\cos \Theta = \frac{r^2 - p^2 - 1}{2p},
$
we can check that the argument  $p \, \omega \sin \Theta  =1$ at $r= r_{\pm}$, where
$$
r_{\pm} \equiv  \sqrt{ p^2 + 1 \pm 2 p \sqrt{1 - \frac{1}{p^2 \omega^2}    }   }.
$$
We proceed to split the integral into three parts where in two of them, we can use \eqref{BesselExpandsmall}.
Denoting the integrand in \eqref{oneloop} to be $F(p,r)$ (and taking note of the respective domains of 
the asymptotic expansion for small or large $z \equiv p\, \omega \sin \Theta $), we write
$I(q,\omega )$ as a sum of three terms as follows. 
\bea
I (q, \omega) &=& \int^{\frac{1}{\omega}}_0 dp\,\, \int^{p+1}_{|p-1|} dr\,\, F(p,r)|_{z<1} +
 \int^{L/q}_{\frac{1}{\omega}} dp\,\, \int^{r_-}_{|p-1|} dr\,\, F(p,r)|_{z<1}  \cr
&\qquad&\qquad \qquad +
\int^{L/q}_{\frac{1}{\omega}} dp\,\, \int^{p+1}_{r_+} dr\,\, F(p,r)|_{z<1}
+ \int^{L/q}_{\frac{1}{\omega}} dp\,\, \int^{r_+}_{r_-} dr\,\, F(p,r)|_{z>1}  \cr
\label{integralS}
&=& \int^{L/q}_0 dp\,\, \int^{p+1}_{|p-1|} dr\,\, F(p,r)|_{z<1} - 
 \int^{L/q}_{\frac{1}{\omega}} dp\,\, \int^{r_+}_{r_-} dr\,\, F(p,r)|_{z<1} 
+ \int^{L/q}_{\frac{1}{\omega}} dp\,\, \int^{r_+}_{r_-} dr\,\, F(p,r)|_{z>1} \cr \cr
&\equiv& I_1 (q,\omega ) + I_2 (q, \omega ) + I_3 (q, \omega ).
\eea
If we take $\omega = \frac{q}{L}$, 
the domains of the virtual momenta integrals in $I_2, I_3$ close up, leaving one with only $I_1$ in \eqref{integralS}.
One is led to consider both the regimes  $0 < \omega \leq \frac{q}{L}$ (with only $I_1$) and $1> \omega \geq \frac{q}{L}$ 
(with $I(q,\omega) = I_1+I_2+I_3$) separately. When implementing the cutoff regularization, we noted
earlier that one should identify $ L/q = \Lambda_{phy} / H $ and the renormalized finite one-loop term is then obtained 
by effectively replacing $q$ with $H(\tau_q)$ in the argument of the logarithmic running. Adopting the same 
procedure here, we see that apart from $I_1$, both $I_2, I_3$ should be added to the one-loop term for 
$1> \omega \geq \frac{H}{\mu}$ 
and both should vanish at 
the transition value $\omega = \frac{H}{\mu}$. For  $0 < \omega \leq \frac{H}{\mu}$, only $I_1$ remains and one should have a smooth 
$\omega = 0$ limit for $I_1$. 

Although it turns out that \eqref{integralS} is difficult to compute for the cases of interest in this paper,  some simplification can be applied for us to explicitly compute the one-loop correction nonetheless. In particular, we found that there is no one-loop contribution arising from $I_3$. 
For computational ease, 
we perform a change
of variables in \eqref{integralS} by defining
$$
l = \frac{1}{p\omega \sin \Theta }, \qquad  s = \frac{1}{p \omega},
$$
after which one obtains
\bea
\label{oneloop3}
I_3(q,\omega) &=&   \int^{L/q}_{\frac{1}{\omega}} dp\, \int^{r_+}_{r_-} dr \, \left[ F_1(p,r) + F_2 (p,r) \right] J_0 (p\, \omega \sin \Theta ) \cr
&=& \frac{1}{\omega^2} \int^1_{\frac{q}{L\omega}}ds\, \frac{1}{s}\, \int^1_s \frac{dl}{l^2 \sqrt{l^2 - s^2}} \left[
\frac{1}{r_{+}(s,l)} F (s, r_+ (s,l)) + \frac{1}{r_{-}(s,l)} F (s, r_- (s,l))
\right], \nonumber \\
\eea
where we have defined
\bea
r_{\pm}(s, l) &\equiv&  \sqrt{ \left(  \frac{1}{s \omega} \right)^2 + 1 \pm 2 \left(  \frac{1}{s \omega} \right) \sqrt{1 - \frac{s^2}{ l^2}    }   }, \cr
F(s, r_{\pm} (s,l)) &\equiv& \left[  F_1(s,r_\pm (s,l)) + F_2 (s,r_\pm (s,l) )  \right] J_0 ( l ).
\eea
The limiting radii $r_\pm$ which mark the boundary of the previous $r$-integral  in \eqref{oneloop3} are now degenerate at $l=1$, with $r_\pm = r_\pm (s, 1)$. As we will see, the integrand has a rapidly oscillatory behavior as we send the cutoff to infinity which is an indeterminate limit.

\subsection{Scalar field}
In the following, we discuss the one-loop correction for the scalar field in detail before presenting the results of the fermions and gauge fields. 
We begin with the integral $I_1$ which we find to be
of the form 
\bea
I_1 (q, \omega ) &=& I_c (q) + 
\frac{\omega^2}{5040} \Bigg[  \tilde{L} ( -18 + 9\tilde{L} + 134 \tilde{L}^2 + 1023 \tilde{L}^3 + 1080 \tilde{L}^4 + 220 \tilde{L}^5 ) \cr
&&\qquad \qquad \qquad \qquad  +18 \text{Log} (\tilde{L} ) - 18 (1+ \tilde{L} )^4 (-1 + 4\tilde{L} - 10 \tilde{L}^2 
+20 \tilde{L}^3 ) \text{Log} (1 + \tilde{L}^{-1} ) \Bigg]  \nonumber \\
\eea
where we define $\tilde{L} = L/q$, and $I_c (q)$ corresponds to the leading order term in \eqref{BesselExpandsmall}, explicitly
\be
I_c (q) = \frac{1}{60} \left[  \tilde{L} ( 8 + 11 \tilde{L} + 6 \tilde{L}^2 + 3 \tilde{L}^3 ) - 8 \text{Log} (\tilde{L}) + 
4 ( -2 -5\tilde{L}^3 + 3\tilde{L}^5 ) \text{Log} (1 + \tilde{L}^{-1} ) \right]. 
\ee
Thus, after discarding terms of the form $\omega^n, n>0$ with coefficients which diverge 
polynomially in $\tilde{L}$, we can read off
\be
\label{I1}
I_1 (q, \omega ) = I_c (q) -\frac{1}{280}\omega^2 \, \text{Log}\, \left( \frac{L}{q} \right) +
\ldots 
\ee
The term containing $ \omega^2$ comes from $k=1$ term in \eqref{BesselExpandsmall}.
All higher $k$-terms yield no logarithmic contributions to the one-loop correction. 
After renormalization to take into account terms that diverge as
$\tilde{L} \rightarrow \infty$, and further replacing $\tilde{L} = \Lambda_{phy}/H(\tau_q)$ as explained earlier, 
the residual logarithmic running reads 
\be
\label{oneI1}
I_1 (q,\omega) = \left( \frac{2}{15} - \frac{1}{280}\omega^2 \right) \text{Log}\, \left( \frac{H}{\mu} \right),
\ee
which smoothly reduces to the commutative one-loop term as $\omega \rightarrow 0$. This is the one-loop correction for $\omega < \frac{H}{\mu}$. For the rest of the domain of $\omega$, the added complication is that one needs to also include $I_2$ and $I_3$, both of which should vanish at $\omega = \frac{H}{\mu}$. 

Before we proceed to compute them, it serves as a good consistency check to study if we also obtain the same one-loop correction $I_1$ via dimensional regularization. A straightforward computation reveals that it yields a one-loop term identical to what we obtained via cutoff method above. Explicitly we find 
\bea
J_1 (q, \omega ) &=& J_c (q) + \frac{\omega^2}{\delta} \int^{\infty}_1 dP \left(
-\frac{1}{6 P^{-\frac{6}{\delta}}} + \frac{1}{60 P^{-\frac{4}{\delta}}} - \frac{1}{24 P^{-\frac{3}{\delta}}} + \frac{1}{40P^{-\frac{2}{\delta}}} - \frac{1}{120 P^{-\frac{1}{\delta}}} + \frac{1}{280}
- \frac{1}{560 P^{\frac{1}{\delta}}} + \ldots 
\right) \cr
\label{J1NC}
&&\,\,\,\,\,\, + \frac{\omega^4}{\delta} \int^{\infty}_1 dP \left( 
\frac{1}{ 120 P^{-\frac{8}{\delta}}} - \frac{1}{168 P^{-\frac{6}{\delta}}} + \frac{1}{480 P^{-\frac{5}{\delta}}}
-\frac{1}{720 P^{-\frac{4}{\delta}}} + \frac{1}{3360 P^{-\frac{3}{\delta}}}
\right) + \mathcal{O} (\omega^6)
\eea
where as introduced in Section 2, in dimensional regularization, we define
$$I_1 (q,\omega) = q^\delta \frac{\pi^{\delta/2}}{\Gamma (1 + \delta/2 )} J_1(q, \omega),$$ 
and in \eqref{J1NC}, $J_c(q)$ refers to the corresponding term in the commutative case. In \eqref{J1NC}, we see that 
the only finite term in the integrand is the numerical constant $-\frac{1}{280}$. The integrands for all other terms higher-order in $\omega$ (which follow from the higher-order terms
in the `$k$'-expansion of \eqref{BesselExpandsmall}) can be obtained straightforwardly and they contain no such finite constant. 
This precisely reproduces the result \eqref{oneI1} after including logarithmic terms that arise
from the analytic continuation of the internal loop wavefunctions and scale factors.

For the second integral $I_2$ (to be added to the one-loop correction for $\omega > H/\mu$), we find correction terms which are non-perturbative in $\omega$. Schematically, we find it to be of the form 
\be
\label{I2}
I_2 (q,\omega ) = \left(  \frac{\mathcal{B}_4}{\omega^4 } + \frac{\mathcal{B}_2}{\omega^2} + \frac{2}{15} - \frac{1}{280}\omega^2 \right) 
\text{Log} \left( \frac{\omega \mu}{H} \right) + \ldots
\ee
where the ellipses stand for
terms which diverge with the cutoff, and the coefficients $\mathcal{B}_4, \mathcal{B}_2$ are finite constants which can be determined up to 
arbitrary order in the `$k$'-expansion of \eqref{BesselExpandsmall}. For example, working up to and including $k=2$ in 
\eqref{BesselExpandsmall}, we find
$\mathcal{B}_4 = \frac{1}{8} - \frac{1}{48} + \frac{1}{1024} +\ldots$, 
$\mathcal{B}_2 = -\frac{1}{4} + \frac{1}{32} -  \frac{1}{768} +\ldots$, with the ellipses representing higher-$k$ terms 
that are increasingly smaller in a converging series. It is important to note that \eqref{I2} vanishes at the transition point 
$\omega = H/\mu$.

Again, we could compute this one-loop term via dimensional regularization. Denoting $I_2 (q,\omega) = q^\delta \frac{\pi^{\delta/2}}{\Gamma (1 + \delta/2 )} J_2(q, \omega)$, for the first few values of $k$, we find 
\bea
&&k=0: \,\, J_2 (q, \omega ) = \frac{1}{\delta} \int^{\infty}_{(1/\omega)^\delta} dP \,\left[
-\frac{1}{ P^{-\frac{4}{\delta}}} + \frac{3+\omega^2}{6\omega^2 P^{-\frac{2}{\delta}}} - \frac{1}{4 P^{-\frac{1}{\delta}}}
+ \left( \frac{2}{15} + \frac{1}{8\omega^4} - \frac{1}{4\omega^2} \right) + \mathcal{O}\left(P^{-\frac{1}{\delta}} \right) \right] \cr
&&k=1: \,\, J_2 (q, \omega ) =  \frac{1}{\delta} \int^{\infty}_{(1/\omega)^\delta} dP \,\Bigg[
\frac{\omega^2}{6 P^{-\frac{6}{\delta}}} - \frac{\omega^2}{60  P^{-\frac{4}{\delta}}} + \frac{\omega^2}{24  P^{-\frac{3}{\delta}}}
- \frac{2\omega^4 + 5}{80 \omega^2 P^{-\frac{2}{\delta}}      } \cr
&&\qquad \qquad \qquad \qquad \qquad \qquad + \frac{\omega^2}{120 P^{-\frac{1}{\delta}}} - 
\frac{70-105\omega^2 + 12 \omega^6}{3360 \omega^4} + \mathcal{O} \left(  P^{-\frac{1}{\delta}} \right) \Bigg]\cr
&&k=2: \,\, J_2 (q, \omega ) =  \frac{1}{\delta} \int^{\infty}_{(1/\omega)^\delta} dP \,\Bigg[
-\frac{\omega^4}{120  P^{-\frac{8}{\delta}}} 
+\frac{\omega^4}{1680  P^{-\frac{6}{\delta}}} 
- \frac{\omega^4}{480 P^{-\frac{5}{\delta}}} 
+ \frac{\omega^4}{ 720 P^{-\frac{4}{\delta}}} \cr
\label{J2i}
&&\qquad \qquad \qquad \qquad \qquad \qquad \qquad \qquad
- \frac{\omega^4}{3360 P^{-\frac{3}{\delta}}} 
+ \frac{1}{384 \omega^2 P^{-\frac{2}{\delta}}} 
 +\left(  \frac{1}{1024 \omega^4} - \frac{ 1}{768 \omega^2} \right)
+ \mathcal{O} \left(  P^{-\frac{1}{\delta}} \right)
\Bigg] \nonumber\\
\eea
The one-loop term can again be read off from the finite $P$-independent piece. A crucial 
difference from $J_1 (q,\omega)$ lies in the lower integration limit which yields for every $P$-independent term $C_0$ in the integrand
of \eqref{J2i} a factor of 
\be
\label{CO}
-\frac{C_0}{\delta} \times \left( \omega^{-\delta} \right) \sim -\frac{C_0}{\delta} \left( 1 - \delta \text{Log}(\omega) \right) + \ldots
\ee
As elaborated earlier, after taking into account the analytic continuation of the scale factor and wavefunctions, one further obtains
\bea
I_2(q,\omega) &=& \left( 1 + \delta \log \left( \frac{q}{\mu} \right) + \ldots \right) \left(  -\frac{C_0}{\delta } + \ldots \right) \times \left( 1 +\left( \frac{6}{2} -2 \right) \delta \log \left(\frac{H}{q} \right) + \ldots \right) \cr
&=& -C_0 \log \left(  \frac{H}{\mu} \right) + \ldots
\eea
which together with \eqref{CO} implies that the one-loop correction can be read off as 
\be
I_2 (q,\omega) = C_0 \, \text{Log}\left( \frac{\omega \mu}{H} \right)=
\left(  \frac{\mathcal{B}_4}{\omega^4 } + \frac{\mathcal{B}_2}{\omega^2} + \frac{2}{15} - \frac{1}{280}\omega^2 \right) 
\text{Log} \left( \frac{\omega \mu}{H} \right).
\ee
This agrees precisely with \eqref{I2}. Thus, for the domain $\omega \geq H/\mu$, the sum of both $I_1$ and $I_2$ gives the 
one-loop correction term  
\be
\label{Ionetwo}
I_1 (q,\omega) + I_2 (q,\omega) = \left(  \frac{2}{15} - \frac{1}{280}\omega^2 \right) \text{Log} \left( \omega \right) + \left(  \frac{\mathcal{B}_4}{\omega^4 } + \frac{\mathcal{B}_2}{\omega^2}   \right) 
\text{Log} \left( \frac{\omega \mu}{H} \right).
\ee
Finally, let us turn to the remaining term $I_3$. In \eqref{oneloop3}, we have
\be
\label{Fterms}
F_1 (s, r(s,l)) = \frac{r^2 (s,l) J_0 (l)}{s\omega (1 + s\omega (r(s,l)+1))},\,\,\,
F_2 (s, r(s,l)) = \frac{r^2 (s,l) J_0 (l)}{ (1 + s\omega (r(s,l)+1))^2}.
\ee
To proceed, for computational convenience, one can express the integrand as a Taylor series in $\omega$. The leading term in the large argument expansion of the Bessel function in \eqref{BesselExpandlarge} yields 
\be
\label{I3cut}
 I_3(q,\omega) = \sum^\infty_{n=-4} \sum^{F_n}_{m=1} 
\int^1_{\frac{q}{L\omega}} ds \int^1_sdl\,\,
\frac{s^{n+1}}{l^{\frac{3}{2}}} C_m \left(  \frac{s}{l}  \right)^{2m-2}
\frac{l\sin \left( \frac{1}{l} - \frac{\pi}{4} \right) + 8 \sin \left( \frac{1}{l} + \frac{\pi}{4} \right)}{\sqrt{l^2 - s^2}} \omega^n
\ee
where $C_m$ are constants and 
$$
F_n =
\begin{cases}
\frac{n}{2} + 3, \qquad \text{for even}\,\, n\, \\
\frac{n+1}{2} + 1, \,\,\,\,\,\, \text{for odd}\,\, n\, \\
\end{cases}
$$
At this point unfortunately, we were unable to proceed analytically and hence we resort to a numerical study of the oscillatory integral in \eqref{I3cut} in which we found that the integral is rapidly oscillatory and approaches an indeterminate limit as $q/L\omega \rightarrow 0$, a behavior which persists at other subleading terms in \eqref{BesselExpandlarge}. 
This indicates that there is no one-loop correction term in $I_3$.

Again, it is instructive to approach the calculation via dimensional regularization which, as in the case of $I_2$, translates into 
the existence of a finite constant term in the integrand of the $p-$integral after integrating out the other internal momentum. 
In our new set of coordinates, defining $I_3 (q,\omega) = q^\delta \frac{\pi^{\delta/2}}{\Gamma (1 + \delta/2 )} J_3(q, \omega)$,
\bea
J_3 (q, \omega ) &=& \frac{1}{\delta} \int^{\tilde{L}^\delta}_{(1/\omega)^\delta} dP \,\Bigg[   \sum^\infty_{k=0} \sum^\infty_{n=-4} \sum^{F_n}_{m=1} 
\int^1_s dl\,\, (-1)^k  \cr
&\qquad \qquad & \times  l^{2k}
\frac{s^{n+2}}{l^{\frac{3}{2}}} C_m \left(  \frac{s}{l}  \right)^{2m-2}
\frac{la_{2k} \sin \left( \frac{1}{l} - \frac{\pi}{4} \right) - a_{2k+1}  \sin \left( \frac{1}{l} + \frac{\pi}{4} \right)}{\sqrt{l^2 - s^2}} \omega^n
\Bigg]
\eea
where $s = 1/ p \omega = P^{-\frac{1}{\delta}}/\omega$. We find that there is no finite constant term in the integrand when expressed as a series in 
$s$ (or $P^{-\frac{1}{\delta}}$). One way to see this concretely is to write the integrand formally as an infinite series in $(l-s)$ (which is positive and smaller than unity) so as to develop a series expression for the $l-$integral. Further expanding in $P^{-\frac{1}{\delta}}$ reveals
that all terms are accompanied by either $\sin \left( \frac{1}{s} \right)$ or $\cos \left( \frac{1}{s} \right)$ both of which harbors an essential singularity at $s=0$. Thus, we deduce that there is no one-loop correction arising from $I_3$. For the scalar field then,
taking into account \eqref{Ionetwo} and \eqref{oneI1}, the one-loop correction is
\be
\label{sca}
I(q,\omega)=
\begin{cases}
\left(  \frac{2}{15} - \frac{1}{280}\omega^2 \right) \text{Log}\, \left( \frac{H}{\mu} \right), \, \omega \leq \frac{H}{\mu}  \\
 \left( \frac{2}{15}  - \frac{1}{280}\omega^2 \right)
\text{Log} \left( \omega \right) + \left(  \frac{\mathcal{B}_4}{\omega^4 } + \frac{\mathcal{B}_2}{\omega^2}  \right) 
\text{Log} \left( \frac{\omega \mu}{H} \right), \, \omega > \frac{H}{\mu} \\
\end{cases}
\ee

\begin{figure}[h]
\centering
\includegraphics[width=125mm]{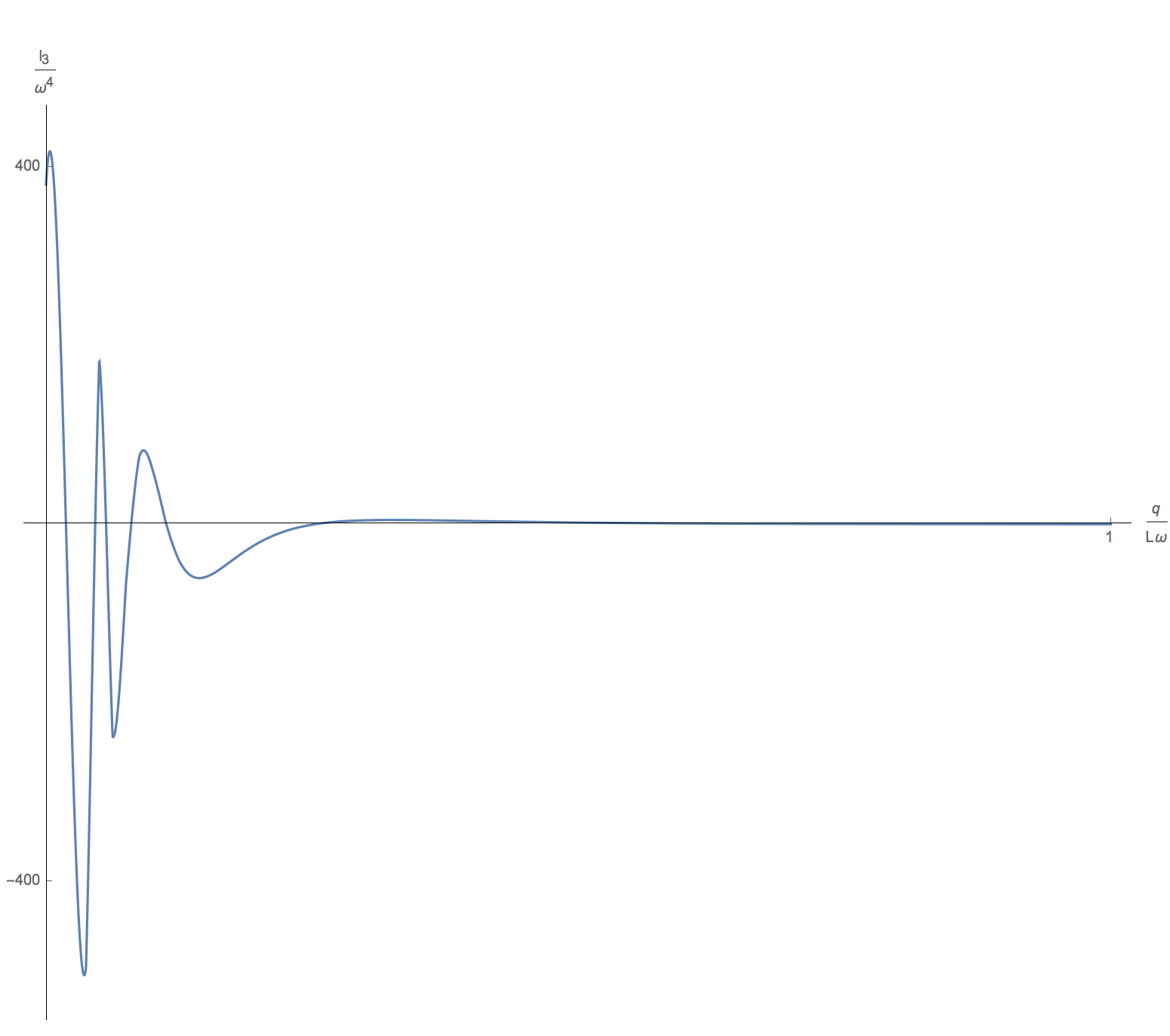}
\caption{A plot of the cofficient of $1/\omega^4$ vs $q/(L\omega)$, with $k=0$ in \eqref{I3cut}, with an interpolating function connecting some set of points obtained by numerical integration for \eqref{I3cut}. We find the same qualitative behavior, in particular the rapidly oscillating behavior at small  $q/(L\omega)$ for other values of $k$ and for other $C_n, \, n>0$. The coefficient asymptotes to zero in the commutative limit $\omega=q/L$, while it becomes rapidly oscillatory as $q/(L\omega) \rightarrow 0$ which is an indeterminate limit. We find similar behavior for the cases of fermion and vector fields.}
\label{fig1}
\end{figure}

\subsection{Fermions and Gauge fields}

The computation of the one-loop correction proceeds similarly for the fermions and gauge fields on the noncommutative spacetime, so we will briefly present the main results. Similar to the scalar field case, we find that while $I_3$ does not contain any finite one-loop terms, both $I_1$ and $I_2$ contribute to the one-loop correction. The following is a compact presentation of various relevant expressions (in both regularization schemes) and the final expression for the one-loop term in each case. 
\begin{itemize}
\item{\textbf{Fermions:}} 

\bea
&&I_1 (q, \omega) = \frac{2}{15}\text{Log} \left(  \frac{L}{q} \right) + \frac{1}{105} \omega^2 \text{Log} \left(  \frac{L}{q} \right), \,\,\,
I_2 (q,\omega ) =-\frac{2}{15}\text{Log} \left( \frac{L\omega}{q} \right) - \frac{1}{105} \omega^2 \text{Log} \left( 
\frac{L\omega}{q} \right)  \cr
&& J_1 (q,\omega )=
\frac{1}{\delta} \int^\infty_{(1/\omega )^\delta} dP
\Bigg[ -\frac{2\omega^2}{105 P^{-\frac{2}{\delta}}} + \frac{1}{105} (14+\omega^2 ) - \frac{\omega^2 P^{\frac{1}{\delta}}}{210}
+ \frac{\omega^2 - 6}{315} P^{-\frac{2}{\delta}} \cr
&&\qquad \qquad \qquad \qquad \qquad
+ \left(  \frac{1}{30} - \frac{\omega^2}{420} \right) P^{-\frac{3}{\delta}} + \mathcal{O} (P^{-\frac{4}{\delta}} ) \Bigg] \cr
&&J_2 (q,\omega ) = \frac{1}{\delta} \int^\infty_{(1/\omega)^\delta} \, dP \left[ 
\frac{\omega^4}{1260P^{-\frac{4}{\delta}}}
+ \left(  -\frac{\omega^4}{2310} + \frac{2\omega^2}{105}    \right) \frac{1}{ P^{-\frac{2}{\delta}}} 
+\frac{\omega^4}{5040 P^{-\frac{1}{\delta}}}
- \left(  \frac{2}{15} + \frac{\omega^2}{105} \right) 
+ \mathcal{O} (P^{-\frac{1}{\delta} } )
\right]\nonumber \\ \cr
\label{fer} 
&&I(q,\omega)=
\begin{cases}
\frac{2}{15}\text{Log} \left(  \frac{\mu}{H} \right) + \frac{1}{105} \omega^2 \text{Log} \left(  \frac{\mu}{H} \right), \, \omega \leq \frac{H}{\mu}   \\
-\frac{2}{15}\text{Log} \left( \omega \right) - \frac{1}{105} \omega^2 \text{Log} \left( \omega \right), \, \omega > \frac{H}{\mu}\\
\end{cases}
\eea
\item{\textbf{Gauge fields:}}

\bea
&&I_1 (q,\omega ) = \frac{1}{15} \text{Log} \left(  \frac{L}{q} \right) + \frac{1}{140} \omega^2 \text{Log} \left(  \frac{L}{q} \right), \,\,\,
I_2 (q,\omega ) = -\frac{1}{15}\text{Log} \left( \frac{L \omega}{q} \right) + \frac{\omega^2}{140} \text{Log} \left(  \frac{q}{L\omega} \right) \cr
&&J_1 (q,\omega ) = \frac{1}{\delta} \int^\infty_{(1/\omega)^\delta} dP
\Bigg[
-\frac{\omega^2}{70 P^{-\frac{2}{\delta}}} + \left( \frac{1}{15} + \frac{\omega^2}{140} \right) 
- \frac{\omega^2 P^{-\frac{1}{\delta}}}{280}
+ \left( -\frac{1}{70} + \frac{\omega^2}{504} \right) P^{-\frac{2}{\delta}} \cr
&& \qquad \qquad \qquad \qquad \qquad 
+ \left( \frac{1}{60} - \frac{\omega^2}{840} \right) P^{-\frac{3}{\delta}} 
+ \mathcal{O} (P^{-\frac{4}{\delta}} )
\Bigg]\cr
&&J_2 (q,\omega ) = \frac{1}{\delta} \int^\infty_{(1/\omega)^\delta} dP
\left[  -\frac{\omega^4}{1260 P^{-\frac{4}{\delta}}}   + \left(  \frac{\omega^4}{1848} + \frac{\omega^2}{70}         \right)    \frac{1}{P^{-\frac{2}{\delta}}} 
-\frac{\omega^4}{5040 P^{-\frac{1}{\delta}}}
- \left( \frac{1}{15} + \frac{\omega^2}{140} \right)
+ \mathcal{O}(P^{-\frac{1}{\delta}})         \right] \nonumber \\ \cr 
\label{Gau}
&&I(q,\omega)=
\begin{cases}
\frac{1}{15}\text{Log} \left(  \frac{\mu}{H} \right) + \frac{1}{140} \omega^2 \text{Log} \left(  \frac{\mu}{H} \right), \, \omega \leq \frac{H}{\mu}   \\
-\frac{1}{15}\text{Log} \left( \omega \right) - \frac{1}{140} \omega^2 \text{Log} \left( \omega \right), \, \omega > \frac{H}{\mu}\\
\end{cases}
\eea
\end{itemize}
We note that both terms in $I_2$ can be attributed just to the first two terms in \eqref{BesselExpandlarge}, with the $\omega^2$ term arising from the subleading ($k=1$) terms in the `$k$'-expansion in \eqref{BesselExpandlarge}. 
All other $k \geq 2$ do not contribute to the one-loop logarithmic running as we 
observe in both independent methods of regularization. In both \eqref{fer} and 
\eqref{Gau}, we display only the most singular terms arising from $k=0,1,2$. 
Evidently, the fermions and gauge fields tend to admit a similar form of one-loop running. For $\omega \leq  H/\mu$,
the correction term goes as $\sim (C_1 + C_2\, \omega^2 ) \text{Log} (\frac{H}{\mu})$ for some constants $C_1, C_2
$. This is also the case for the scalar field but with an opposite sign attached to the $\omega^2$ term. 

For $\omega \geq H/\mu$, we simply replace $H/\mu$ with $\omega$ in the logarithmic argument for the fermions and gauge fields,
whereas for the scalar field, there is an additional term $\left( \frac{\mathcal{B}_4}{\omega^4} + \frac{\mathcal{B}_2}{\omega^2} \right) 
\text{Log} \left( \frac{\omega \mu}{H} \right)$ which goes to zero at $\omega = H/\mu$ ensuring continuity. This additional term however
implies that for $\omega > H/ \mu$, the one-loop suddenly increases once $\omega$ goes beyond this value if $H/\mu$ is small. This feature is absent for the fermions and gauge fields. 

For all massless fields, there is a kink in the one-loop correction at this critical value equivalent 
to $q = \sqrt{   \frac{H}{\theta \mu} }$. This `critical value' of $q$ beyond which $I_2$ contributes to the one-loop correction 
has an interesting interpretation in terms of UV/IR mixing. As explained in for example \cite{Minwalla}, if a field
$\phi$ is non-vanishing over a small spatial region of size $\delta \ll \sqrt{\theta}$, then $\phi \star \phi$ is non-vanishing over a much larger 
region of size $\theta/\delta$. This `nonlocality' due to the noncommutative deformation implies that the IR dynamics receive
contributions from high-energy virtual particles. For the non-planar graph, a virtual particle of energy $E \gg 1/\sqrt{\theta}$ will upon 
interaction produce measurable effects at energy $1/(\theta E)$. In particular, we can relate UV and IR effective cutoffs roughly as
$$
\Lambda_{IR} \sim \frac{1}{\theta \Lambda_{UV} }.
$$
In our evaluation of the one-loop correction via cutoff regularization, we imposed the UV cutoff $L = q \Lambda_{phy}/H$, and the one-loop logarithmic running has $\Lambda_{phy}/H$ 
as its argument (after renormalization, we replace $\Lambda_{phy} \rightarrow \mu$).  For $q > \sqrt{   \frac{H}{\theta \mu} }$, the term $I_2$ yields
a logarithmic running of the form $\text{Log} (H/( q^2 \theta \mu))$, compatible with a general expectation that interaction on the noncommutative background induces
physical effects at the scale $H/ (\theta q \Lambda_{phy})$. For the fermion and vector fields, this translates to replacing 
\be
\frac{\mu}{H} \rightarrow \frac{H}{\omega \mu}
\ee
in the logarithmic argument with the prefactor $C_1 + C_2 \omega^2$ unchanged which, together with the $I_1$, implies that we have a logarithmic running that goes as  
$\text{Log} (\omega )$. 
For the scalar field, apart from such a replacement, we also have 
an additional term $\left( \frac{\mathcal{B}_4}{\omega^4} + \frac{\mathcal{B}_2}{\omega^2} \right) 
\text{Log} \left( \frac{\omega \mu}{H} \right)$ 
giving a discontinuity in gradient that grows with decreasing $\omega_{crit} = H/\mu$. Modes of superhorizon scales enter into the one-loop logarithmic term spoiling the scale invariance of the term in the commutative case. We note however that $\omega$ is invariant under the rescaling symmetry 
$$a \rightarrow \lambda a, x\rightarrow x/\lambda, q\rightarrow \lambda q.$$ 
Under this symmetry, $\theta \rightarrow \theta / \lambda^2 $, but the `proper' noncommutative parameter $a^2 \theta$ (that arises from the commutator of proper instead of comoving coordinates) is invariant. 

The limit $\omega =1$ corresponds to the point where the coordinates' uncertainty relation is saturated. For the fermions and gauge fields, 
we find that this precisely corresponds to the point where the one-loop running vanishes, with the one-loop correction being negative for $\omega>1$ and positive otherwise. For the scalars, it appears that this point is close to a local minimum instead. Although we have not explicitly imposed this noncommutativity-related UV cutoff on $q$, it is not clear whether our one-loop computation 
will persist to be reliable in this regime, since this probes length scale smaller than the noncommutative scale. On the other hand, 
the $\omega = 0$ limit is smooth, compatible with the limiting behavior of the Bessel function
in \eqref{oneloop2}. This is not always the case for standard QFT with an S-matrix defined on a non-commutative background where, as explained in 
\cite{susskind}, UV/IR mixing can sometimes lead to non-analytic behavior leading to a singular $\theta = 0$ limit.

\begin{figure}
    \centering
    \begin{subfigure}[b]{0.31\textwidth}
        \includegraphics[width=\textwidth]{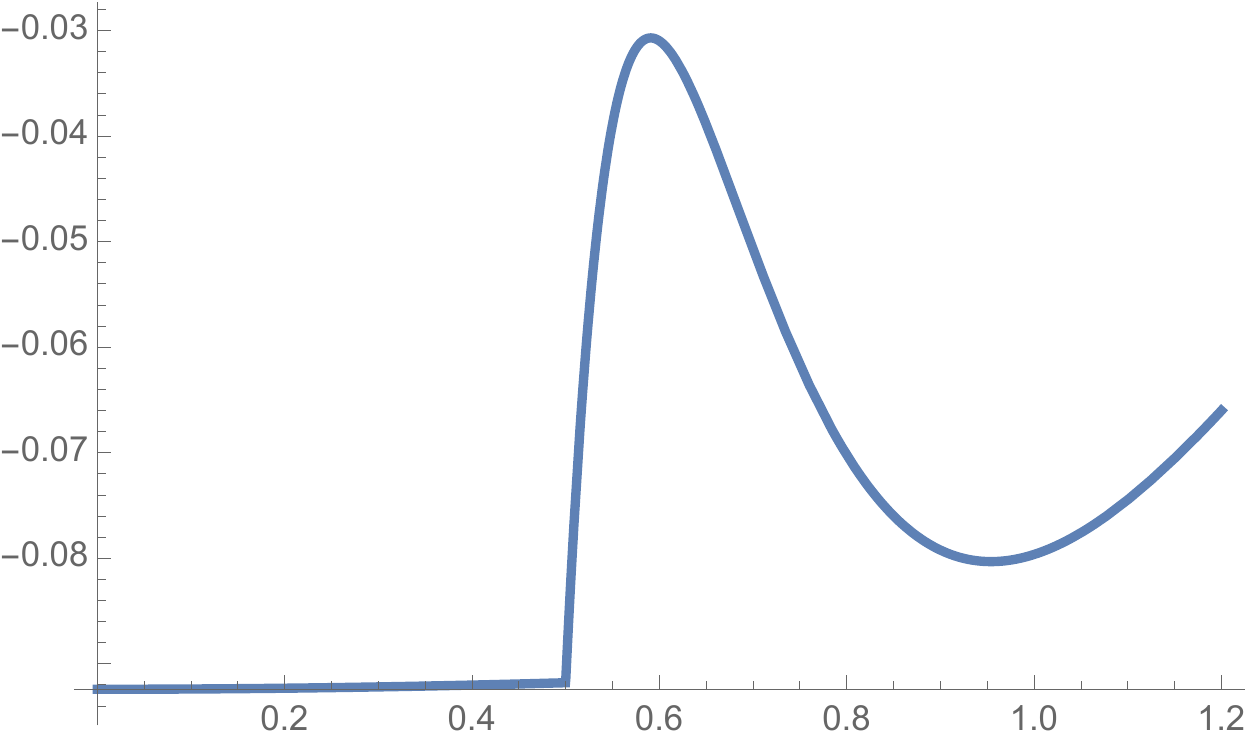}
        \caption{Scalar}
        \label{fig:scalar}
    \end{subfigure}
    ~ 
    \begin{subfigure}[b]{0.31\textwidth}
        \includegraphics[width=\textwidth]{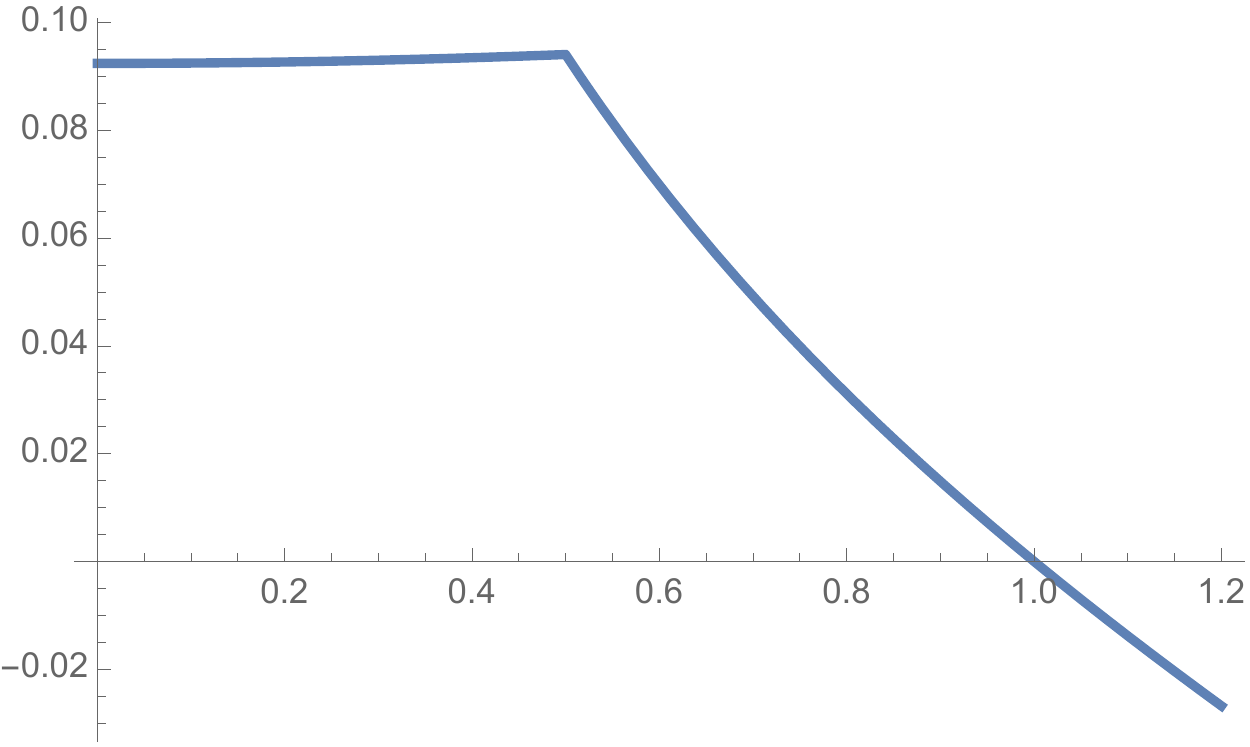}
        \caption{Fermion}
        \label{fig:fermion}
    \end{subfigure}
    ~ 
    \begin{subfigure}[b]{0.31\textwidth}
        \includegraphics[width=\textwidth]{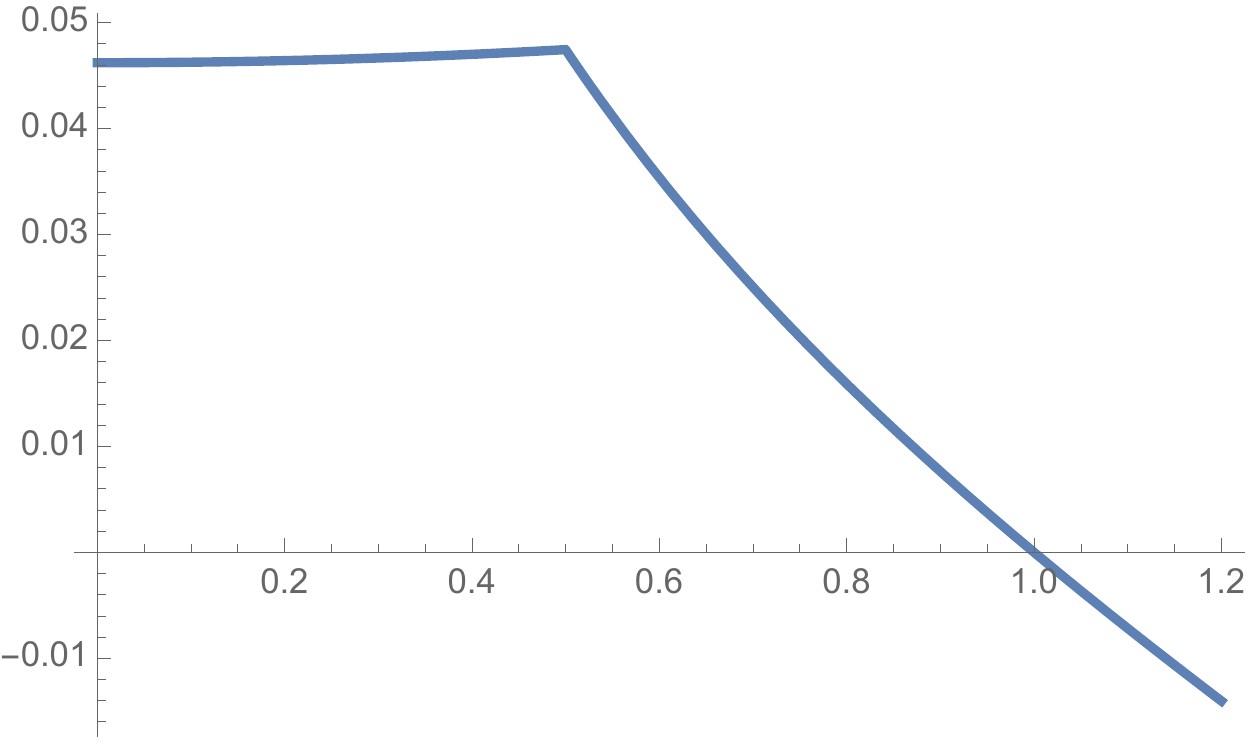}
        \caption{Gauge field}
        \label{fig:gauge}
    \end{subfigure}
    \caption{Plots of one-loop correction $I(q,\omega)$ of various massless fields vs $\omega=q^2\theta$ for the choice of $H/\mu = 0.5$. For all cases, the first derivative is discontinuous at the transition value $\omega = H/\mu$. The fermions and gauge fields display similar shape dependence, and their one-loop corrections vanish at $\omega =1$ after which each changes in sign.  }
\label{fig:masslessfields}
\end{figure}


\section{Conclusion}

After performing a non-commutative deformation of the trilinear interaction Hamiltonians that pertain to various forms of massless spectator matter 
fields interacting gravitationally with the inflaton, we have derived the one-loop correction to the primordial curvature two-point function, in 
particular those that arise in the non-planar diagrams. The fact that they are momentum-dependent indicates that the vacuum fluctuations of the 
energy-momentum tensor source that of the curvature fluctuation even for distances beyond horizon scales - a clear signature of non-locality. 

This effect of noncommutative spacetime geometry on one-loop corrections
was studied based on the simplest noncommutative relation between comoving coordinates which 
preserves the usual classical FRW evolution. The noncommutative deformation was performed via deforming the trilinear interaction 
Hamiltonian term (in the usual commutative setting) by equipping the algebra of functions on the inflationary background with a 
spatially noncommutative Moyal star product. For all massless fields, there is a kink in the one-loop correction at this critical value equivalent to $q = \sqrt{   \frac{H}{\theta \mu} }$. This critical value of $q$ carries a plausible interpretation in terms of the phenomenon of UV/IR mixing. A logarithmic term of the form 
$\text{Log} \left(  \frac{\omega \mu}{H} \right)$ appears for momentum modes above the critical value, capturing the non-local
spreading of virtual particles which are running in the loop and having energy close to the UV cutoff. For the fermions and gauge field, 
this term combines neatly with the original logarithmic term to yield the overall logarithmic running $\text{Log} \left( \omega \right)$. This
holds for the scalar field as well but in this case, there is an additional term of the form 
$\left( \frac{\mathcal{B}_4}{\omega^4} + \frac{\mathcal{B}_2}{\omega^2} \right) 
\text{Log} \left( \frac{\omega \mu}{H} \right)$, an effect which vanishes at the critical value. 

Another interesting point is $\omega =1$ which, roughly speaking,
corresponds to the point where the coordinates' uncertainty relation (due to the noncommutative deformation) is saturated. For the fermions and gauge fields, 
we find that this remarkably corresponds to the point where the one-loop running vanishes, with the one-loop correction being negative for $\omega>1$ and positive otherwise. For the scalars, it appears that this point is close to a local minimum instead. Although we have not explicitly imposed this noncommutativity-related UV cutoff on $q$, it is not clear whether our one-loop computation will persist to be reliable in this regime, since this probes length scale smaller than the noncommutative scale. It would be interesting to see if the vanishing of the one-loop correction persists beyond one-loop order for the fermions and gauge fields. 

Our computation of the one-loop correction was performed via two distinct regularization methods - (i)imposing a momentum cutoff (ii)dimensional regularization including that of modes' wavefunctions. In all cases, we verified that both methods which yield different infinite terms to be absorbed by renormalization give
nevertheless the same finite one-loop term, thus completing a consistency cross-check. In the process, we also refined the accuracy of some of the one-loop terms presented in the usual commutative setting reported previously. We also found that the $\omega = 0$ limit is smooth, compatible with the fact that we have performed the noncommutative deformation under the same assumption. The one-loop corrections break spatial isotropy by being functions of spatial components of the matrix-valued noncommutative parameter and furnish an example of how in principle, a quantum gravitational effect could introduce  signatures of non-locality and anisotropy in loop corrections of the primordial power spectrum computed in the Schwinger-Keldysh formalism.

\section*{Acknowledgments}

I am grateful to Neal Snyderman, Ori Ganor and Eugene Lim for useful discussions. This work is supported by 
a research fellowship given by the School of Physical and Mathematical Sciences, Nanyang Technological University of Singapore.


\begin{thebibliography}{10}



\bibitem{Snyder} 
  H.~S.~Snyder,
  ``Quantized space-time,''
  Phys.\ Rev.\  {\bf 71}, 38 (1947).
  doi:10.1103/PhysRev.71.38

\bibitem{Yang} 
  C.~N.~Yang,
  ``On quantized space-time,''
  Phys.\ Rev.\  {\bf 72}, 874 (1947).
  doi:10.1103/PhysRev.72.874

\bibitem{Myers} 
  R.~C.~Myers,
  ``Dielectric branes,''
  JHEP {\bf 9912}, 022 (1999)
  doi:10.1088/1126-6708/1999/12/022
  [hep-th/9910053].


\bibitem{Witten} 
  N.~Seiberg and E.~Witten,
  ``String theory and noncommutative geometry,''
  JHEP {\bf 9909}, 032 (1999)
  doi:10.1088/1126-6708/1999/09/032
  [hep-th/9908142].

\bibitem{Henry} 
 S.-H.~Henry Tye,
  ``Brane inflation: String theory viewed from the cosmos,''
  Lect.\ Notes Phys.\  {\bf 737}, 949 (2008)
  [hep-th/0610221].

\bibitem{Shiu} 
  C.~S.~Chu, B.~R.~Greene and G.~Shiu,
  ``Remarks on inflation and noncommutative geometry,''
  Mod.\ Phys.\ Lett.\ A {\bf 16}, 2231 (2001)
  doi:10.1142/S0217732301005680
  [hep-th/0011241].

\bibitem{Buchel} 
  A.~Buchel and A.~Ghodsi,
  ``Braneworld inflation,''
  Phys.\ Rev.\ D {\bf 70}, 126008 (2004)
  doi:10.1103/PhysRevD.70.126008
  [hep-th/0404151].


\bibitem{Alexander} 
  S.~Alexander, R.~Brandenberger and J.~Magueijo,
  ``Noncommutative inflation,''
  Phys.\ Rev.\ D {\bf 67}, 081301 (2003)
  doi:10.1103/PhysRevD.67.081301
  [hep-th/0108190].

\bibitem{Weinberg1} 
  S.~Weinberg,
  ``Quantum contributions to cosmological correlations,''
  Phys.\ Rev.\ D {\bf 72}, 043514 (2005)
  doi:10.1103/PhysRevD.72.043514
  [hep-th/0506236].

\bibitem{Senatore} 
  L.~Senatore and M.~Zaldarriaga,
  ``On Loops in Inflation,''
  JHEP {\bf 1012}, 008 (2010)
  doi:10.1007/JHEP12(2010)008
  [arXiv:0912.2734 [hep-th]].

\bibitem{Eugene} 
  P.~Adshead, R.~Easther and E.~A.~Lim,
  ``Cosmology With Many Light Scalar Fields: Stochastic Inflation and Loop Corrections,''
  Phys.\ Rev.\ D {\bf 79}, 063504 (2009)
  doi:10.1103/PhysRevD.79.063504
  [arXiv:0809.4008 [hep-th]].

\bibitem{Chai} 
  K.~Chaicherdsakul,
  ``Quantum Cosmological Correlations in an Inflating Universe: Can fermion and gauge fields loops give a scale free spectrum?,''
  Phys.\ Rev.\ D {\bf 75}, 063522 (2007)
  doi:10.1103/PhysRevD.75.063522
  [hep-th/0611352].

\bibitem{Tanaka} 
  T.~Tanaka and Y.~Urakawa,
  ``Loops in inflationary correlation functions,''
  Class.\ Quant.\ Grav.\  {\bf 30}, 233001 (2013)
  doi:10.1088/0264-9381/30/23/233001
  [arXiv:1306.4461 [hep-th]].

\bibitem{Giddings} 
  S.~B.~Giddings and M.~S.~Sloth,
  ``Semiclassical relations and IR effects in de Sitter and slow-roll space-times,''
  JCAP {\bf 1101}, 023 (2011)
  doi:10.1088/1475-7516/2011/01/023
  [arXiv:1005.1056 [hep-th]].


\bibitem{Seery1} 
  D.~Seery,
  ``One-loop corrections to a scalar field during inflation,''
  JCAP {\bf 0711}, 025 (2007)
  doi:10.1088/1475-7516/2007/11/025
  [arXiv:0707.3377 [astro-ph]].

\bibitem{Feng} 
  K.~Feng, Y.~F.~Cai and Y.~S.~Piao,
  ``IR Divergence in Inflationary Tensor Perturbations from Fermion Loops,''
  Phys.\ Rev.\ D {\bf 86}, 103515 (2012)
  doi:10.1103/PhysRevD.86.103515
  [arXiv:1207.4405 [hep-th]].

\bibitem{Akofor1} 
  E.~Akofor, A.~P.~Balachandran, S.~G.~Jo, A.~Joseph and B.~A.~Qureshi,
  ``Direction-Dependent CMB Power Spectrum and Statistical Anisotropy from Noncommutative Geometry,''
  JHEP {\bf 0805}, 092 (2008)
  doi:10.1088/1126-6708/2008/05/092
  [arXiv:0710.5897 [astro-ph]].


\bibitem{Akofor2} 
  E.~Akofor, A.~P.~Balachandran, A.~Joseph, L.~Pekowsky and B.~A.~Qureshi,
  ``Constraints from CMB on Spacetime Noncommutativity and Causality Violation,''
  Phys.\ Rev.\ D {\bf 79}, 063004 (2009)
  doi:10.1103/PhysRevD.79.063004
  [arXiv:0806.2458 [astro-ph]].

\bibitem{Nautiyal}
  A.~Nautiyal,
  ``Anisotropic non-gaussianity with noncommutative spacetime,''
  Phys.\ Lett.\ B {\bf 728} (2014) 472
  doi:10.1016/j.physletb.2013.12.007
  [arXiv:1303.4159 [astro-ph.CO]].


\bibitem{Chen} 
  X.~Chen, Y.~Wang and Z.~Z.~Xianyu,
  ``Loop Corrections to Standard Model Fields in Inflation,''
  JHEP {\bf 1608}, 051 (2016)
  doi:10.1007/JHEP08(2016)051
  [arXiv:1604.07841 [hep-th]].

\bibitem{Minwalla} 
  S.~Minwalla, M.~Van Raamsdonk and N.~Seiberg,
  ``Noncommutative perturbative dynamics,''
  JHEP {\bf 0002}, 020 (2000)
  doi:10.1088/1126-6708/2000/02/020
  [hep-th/9912072].

\bibitem{Ho} 
  R.~Brandenberger and P.~M.~Ho,
  ``Noncommutative space-time, stringy space-time uncertainty principle, and density fluctuations,''
  Phys.\ Rev.\ D {\bf 66}, 023517 (2002)
  [AAPPS Bull.\  {\bf 12}, no. 1, 10 (2002)]
  doi:10.1103/PhysRevD.66.023517
  [hep-th/0203119].



\bibitem{Cai} 
  R.~G.~Cai,
  ``A Note on curvature fluctuation of noncommutative inflation,''
  Phys.\ Lett.\ B {\bf 593}, 1 (2004)
  doi:10.1016/j.physletb.2004.04.078
  [hep-th/0403134].


\bibitem{Gomis} 
  J.~Gomis and T.~Mehen,
  ``Space-time noncommutative field theories and unitarity,''
  Nucl.\ Phys.\ B {\bf 591}, 265 (2000)
  doi:10.1016/S0550-3213(00)00525-3
  [hep-th/0005129].

\bibitem{Ganor} 
  D.~W.~Chiou and O.~J.~Ganor,
  ``Noncommutative dipole field theories and unitarity,''
  JHEP {\bf 0403}, 050 (2004)
  doi:10.1088/1126-6708/2004/03/050
  [hep-th/0310233].

\bibitem{Mota} 
  T.~S.~Koivisto and D.~F.~Mota,
  ``CMB statistics in noncommutative inflation,''
  JHEP {\bf 1102}, 061 (2011)
  doi:10.1007/JHEP02(2011)061
  [arXiv:1011.2126 [astro-ph.CO]].

\bibitem{Calmet} 
  X.~Calmet and C.~Fritz,
  ``Inflation on a non-commutative space–time,''
  Phys.\ Lett.\ B {\bf 747}, 406 (2015)
  doi:10.1016/j.physletb.2015.06.033
  [arXiv:1506.04049 [hep-th]].


\bibitem{Palma} 
  G.~A.~Palma and S.~P.~Patil,
  ``UV/IR mode mixing and the CMB,''
  Phys.\ Rev.\ D {\bf 80}, 083010 (2009)
  doi:10.1103/PhysRevD.80.083010
  [arXiv:0906.4727 [hep-th]].

\bibitem{DLMF}
[DLMF] NIST Digital Library of Mathematical Functions. http://dlmf.nist.gov/, Release 1.0.19 of 2018-06-22. 
F. W. J. Olver, A. B. Olde Daalhuis, D. W. Lozier, B. I. Schneider, R. F. Boisvert, C. W. Clark, B. R. Miller, and B. V. Saunders, eds.

\bibitem{susskind} 
  A.~Matusis, L.~Susskind and N.~Toumbas,
  ``The IR / UV connection in the noncommutative gauge theories,''
  JHEP {\bf 0012}, 002 (2000)
  doi:10.1088/1126-6708/2000/12/002
  [hep-th/0002075].

























\end{thebibliography}
\end{document}